\documentclass[twocolumn]{aastex631}

\newcommand{\vcirc}{v_{\rm{circ}}}

\newcommand{\mhalf}{M_{1/2}}
\newcommand{\rhalf}{R_{1/2}}

\newcommand{\mhalo}{M_{\rm{halo}}}

\newcommand{\rvir}{r_{\rm{vir}}}

\newcommand{\mstar}{M_{\rm \star}}
\newcommand{\msun}{\rm \, M_{\odot}}

\newcommand{\lsun}{L_{\odot}}

\newcommand{\myr}{\rm \, Myr}
\newcommand{\kms}{{\rm km \, s}^{-1}}

\newcommand{\lcdm}{$\Lambda$CDM}

\newcommand{\mbar}{m^{\rm p}_{\rm bar}}

\newcommand{\mtenq}{\rm m10q_{30}}

\newcommand{\mtenv}{\rm m10v_{30}}

\newcommand{\mtenvB}{\rm m10v_{30} \, B}

\newcommand{\mnine}{\rm m09_{30}}

\newcommand{\feh}{\rm [Fe/H]}

\usepackage{graphicx}

\shorttitle{How invisible stellar halos bias our understanding of ultra-faint galaxies}
\shortauthors{Wheeler et al.}

\begin{document}

\title{How invisible stellar halos bias our understanding of ultra-faint galaxies}
%\title{The Invisible Stellar Halos of Ultra-Faint Galaxies}

\author[0000-0002-2651-7281]{Coral Wheeler}
\affiliation{Department of Physics and Astronomy, California State Polytechnic University, Pomona, Pomona, CA 91768, USA}

\correspondingauthor{Coral Wheeler}
\email{cwheeler@cpp.edu}

\author[0000-0002-3430-3232]{Jorge Moreno}
\affiliation{Department of Physics and Astronomy, Pomona College, Claremont, CA 91711, USA}
\affiliation{Carnegie Observatories, 813 Santa Barbara St., Pasadena, CA 91101, USA}

\author[0000-0003-1848-5571]{M. Katy Rodriguez Wimberly}
\affiliation{Department of Physics and Astronomy, California State University, San Bernardino, San Bernardino, CA 92407, USA}

\author[0000-0002-5908-737X]{Francisco J. Mercado}
\affiliation{Department of Physics and Astronomy, Pomona College, Claremont, CA 91711, USA}
\affiliation{TAPIR, Mailcode 350-17, California Institute of Technology, Pasadena, CA 91125, USA}

\author[0000-0003-4298-5082]{James S. Bullock}
\affiliation{Center for Cosmology, Department of Physics and Astronomy, University of California, Irvine, CA 92697, USA}
\affiliation{Department of Physics and Astronomy, University of Southern California, Los Angeles, CA 90089, USA}

\author[0000-0002-9604-343X]{Michael Boylan-Kolchin}
\affiliation{Department of Astronomy, The University of Texas at Austin, 2515 Speedway, Stop C1400, Austin, TX 78712-1205, USA}

\author[0000-0003-0965-605X]{Pratik J. Gandhi}
\affiliation{Department of Astronomy, Yale University, New Haven, CT 06520, USA}

\author[0000-0003-3217-5967]{Sarah R. Loebman}
\affiliation{Department of Physics, University of California, Merced, CA 95343, USA/}

\author[0000-0003-3729-1684]{Philip F. Hopkins}
\affiliation{TAPIR, Mailcode 350-17, California Institute of Technology, Pasadena, CA 91125, USA}

\begin{abstract}
    We explore how a realistic surface brightness detection limit of $\mu_V \approx 32.5$ mag arcsec$^{-2}$ for stars at the edges of ultra-faint galaxies affects our ability to infer their underlying properties.  We use a sample of 19 galaxies with stellar masses $\approx 400 - 40,000~{\rm M}_\odot$ simulated with FIRE-2 physics and baryonic mass resolution of $30~M_{\odot}$.  The surface brightness cut leads to smaller sizes, lower stellar masses, and lower stellar velocity dispersions than the values inferred without the cut. However, by imposing this realistic limit, our inferred galaxy properties lie closer to observed populations in the mass-size plane,  better match observed velocity dispersions as a function of stellar mass, and better reproduce derived circular velocities as a function of half-light radius. For the most massive galaxies in our sample, the surface brightness cut leads to higher mean $\rm [Fe/H]$ values, but the increase is not enough to match the observed MZR. Finally, we demonstrate that the common \citet{Wolf2010} mass estimator is less accurate when the surface brightness cut is applied.  For our lowest-mass galaxies, in particular, excluding the low-surface brightness outskirts causes us to overestimate their central dark-matter densities and virial masses.  This suggests that attempts to use mass estimates of ultra-faint galaxies to constrain dark-matter physics or to place constraints on the low-mass threshold of galaxy formation must take into account surface brightness limits or risk significant biases.

\end{abstract}


\section{Introduction}
\label{sec:intro} 

The classical missing satellites problem (MSP) is perhaps one of the most well-studied puzzles in galaxy formation theory. First ``discovered'' in 1999 (see also \citealt{Kauffmann1993}), it represents the discrepancy between the number of low-mass ($M_{\rm halo} \lesssim 10^{10}\msun$) dark matter halos predicted to lie within the virial radius of a Milky Way-mass dark matter-only simulation, and the observed number of low-mass ($M_{\rm \star} \lesssim 10^{6}\msun$) galaxies detected around the Milky Way itself \citep{Klypin1999}. Almost immediately, \citet{Bullock2000} suggested a solution: photoionization heating suppresses gas accretion onto low-mass dark matter halos \citep{Efstathiou1992,Babul1992}, placing an effective mass floor on galaxy formation for the bulk of the Universe's history. Reionization feedback remains the standard explanation of the MSP: observations of the ``classical'' Milky Way satellite galaxies (with $M_{\rm \star} \gtrsim 10^{5}\msun$), where our surveys are observationally complete, are in perfect agreement with extrapolations from abundance matching predictions \citep{Kim2018} with a mass threshold consistent with expectations from reionization \citep{Bullock2017}, and hydrodynamical simulations of Milky Way-mass galaxies with sufficient resolution are able to match the number counts of classical satellites \citep{Wang2015,Sawala2016,Wetzel2016,Jung2024}.
% Currently, the classical MSP appears to be ``solved" \citep{Bullock2000,Wetzel2016}. Observations of the ``classical" Milky Way satellite galaxies with ($M_{\rm \star} \gtrsim 10^{5}\msun$), where our surveys are observationally complete, are in perfect agreement with extrapolations from abundance matching predictions \citep{Bullock2017}. 

However, at the lower mass scale of ultra-faint galaxies (UFDs; $M_{\rm \star} \lesssim 10^{5}\msun$), many questions remain. These galaxies are so faint that their discovery and characterization remains a significant observational challenge. The true number of ultra-faint satellites of the Milky Way remains uncertain, limiting our ability to use them as laboratories for galaxy formation and cosmology. For example, while \lcdm~models can make make robust predictions for the number of low-mass dark matter halos as a function of dark matter mass \citep{Press1974,Sheth1999,Wang2024}, including the contribution from subhalos of low-mass satellites or another low-mass galaxy \citep{Wheeler2015,Munshi2019,Jahn2019}, and constrain the surviving subhalo counts around a Milky Way-mass galaxy \citep{Kelley2019}, it remains unclear (a) how many of these low-mass halos are predicted to form galaxies and (b) how the detectability of these galaxies affects our interpretations of low-mass halo occupation.\footnote{ A sharp cutoff in the halo occupation fraction at $\mhalo = 10^8\,M_\odot$, if it exists, can be statistically detected at $1\sigma$ confidence, provided that future surveys are complete to all satellites with $M_V < 0$ and surface brightness $\mu < 30$ mag arcsec$^{-2}$ \citep{Nadler2024}.}

%We focus here on the second question, whether or not the galaxies can be detected by current surveys and telescopes. If the Milky Way satellite population is distributed similarly to the subhalo population of a Milky Way-mass halo in a cosmological dark matter only simulation, the existence of luminosity bias suggests that we may be failing to detect galaxies at low surface brightness, and that there might be hundreds, if not thousands, of undetected low mass galaxies around the Milky Way \citet{Tollerud2008}. If we assume stellar particles trace the potential of NFW dark matter halos, and have a minimal stellar velocity dispersion, the lowest mass dark matter halos may host galaxies with particularly extended stellar populations \citep{Bullock2010}. If this is true, then current observations may represent merely the high-mass tail of a larger distribution of dark matter halos that form UFDs. \citet{Wheeler2019} predict, using FIRE-2 simulations \citep{Hopkins2018a} with $\mbar = 30 \msun$, that ultra-faint galaxies with $\mstar \lesssim 10^4~\msun$ have half-mass radii $>400\rm~ pc$, leading to effective surface brightnesses lower than the detection limits of current surveys $\mu \lesssim 30-32.5$ mag arcsec$^{-2}$ \citep{Koposov2008,Torrealba2018,Han2022}. Although able to match the more extended observed dwarfs, many simulations have struggled to form objects similar to most compact observed UFDs \citep[$\rhalf \lesssim 30\rm ~pc$; ][]{Wheeler2015,Revaz2023,Agertz2019,Sanati2023}.

We focus here on how the low-surface brightness nature of UFDs may bias our interpretations of their underlying properties. It is well known that the census of UFDs around the Milky Way is severely affected by luminosity bias and surface-brightness detection limits \citep{Tollerud2008,DrlicaWagner2020ApJ...893...47D}. However, the low-surface nature of UFDs may also induce biases in the interpretation of galaxies even after they are detected. For example, at fixed velocity dispersion, the lowest mass dark matter halos will host the lowest surface brightness populations \citep{Bullock2010}. If this is true, then our inability to detect low-surface brightness UFDs may {\em systematically} prevent our ability to detect the lowest mass dark matter halos that host galaxies.  That is, the known population of UFDs may represent merely the high-mass tail of a larger distribution of dark matter halos that form ultra-faints. Results from
\citet{Wheeler2019} suggest that this may be the case.  Specifically, using FIRE-2 simulations \citep{Hopkins2018a} with $\mbar = 30 \msun$,  \citet{Wheeler2019} predict a population of ultra-faint galaxies with $\mstar \lesssim 10^4~\msun$ that have half-mass radii in excess of $400\rm~ pc$ and effective surface brightnesses beyond the detection limits of current surveys, $\mu \gtrsim 30-32.5$ mag arcsec$^{-2}$ \citep{Koposov2008,Torrealba2018,Han2022}. Although able to match the more extended observed low mass galaxies, many simulations have struggled to form objects similar to most compact observed UFDs \citep[$\rhalf \lesssim 30\rm ~pc$; ][]{Wheeler2015,Revaz2023,Agertz2019,Sanati2023}.

In the last several years, extended stellar structures have been found around many low mass galaxies \citep{Chiti2021,Filion_and_Wyse2021,Longeard2022,Sestito2023,Tau2024}. These stellar halos might be the result of mergers with other ultra-faint galaxies \citep[][Rivas et al.\ \textit{in prep}]{Deason2014,Tarumi2021,Goater2024,Querci2025}, or the results of tidal interactions with either the Milky Way \citep{Johnston1995,Okamoto2012}, or the LMC/SMC \citep{Sacchi2021}. The lower central surface brightness of ultra-faint galaxies \citep{Simon2019} may make stars at the outskirts more difficult to detect due to their effective surface brightness falling below the background. By applying a simple galaxy ``edge" cut at some fixed fraction of the dark matter halo virial radius (e.g. $15\% ~\rvir$), or by failing to take potential surface brightness effects into account, simulators may be including extended stellar populations that are missing in observed objects. If the observations are restricted to the bright central cores of what are now, or were in the past, more extended stellar objects, could some of the predictions made from these simulations be incorrect?
In this paper, we use the same set of $\mbar = 30\msun$ GIZMO/FIRE-2 simulations of isolated low mass galaxies analyzed in \citet{Wheeler2019}, and investigate what happens to predictions made with these simulations when the stellar halos are included versus when they are excluded. We briefly summarize the simulations in Section \ref{sec:sims}, provide an overview of the results in Section \ref{sec:results}. We discuss our results compared to other work and conclude in Section \ref{sec:conclusion}.

        %%%%%%%%%%%%%%%%%%%%%%%%%%%%%%%%%%%%%%%%%%%%%%%%%%%%%%%%%%%%%% 
	\begin{figure}
		\includegraphics[width=0.48\textwidth]{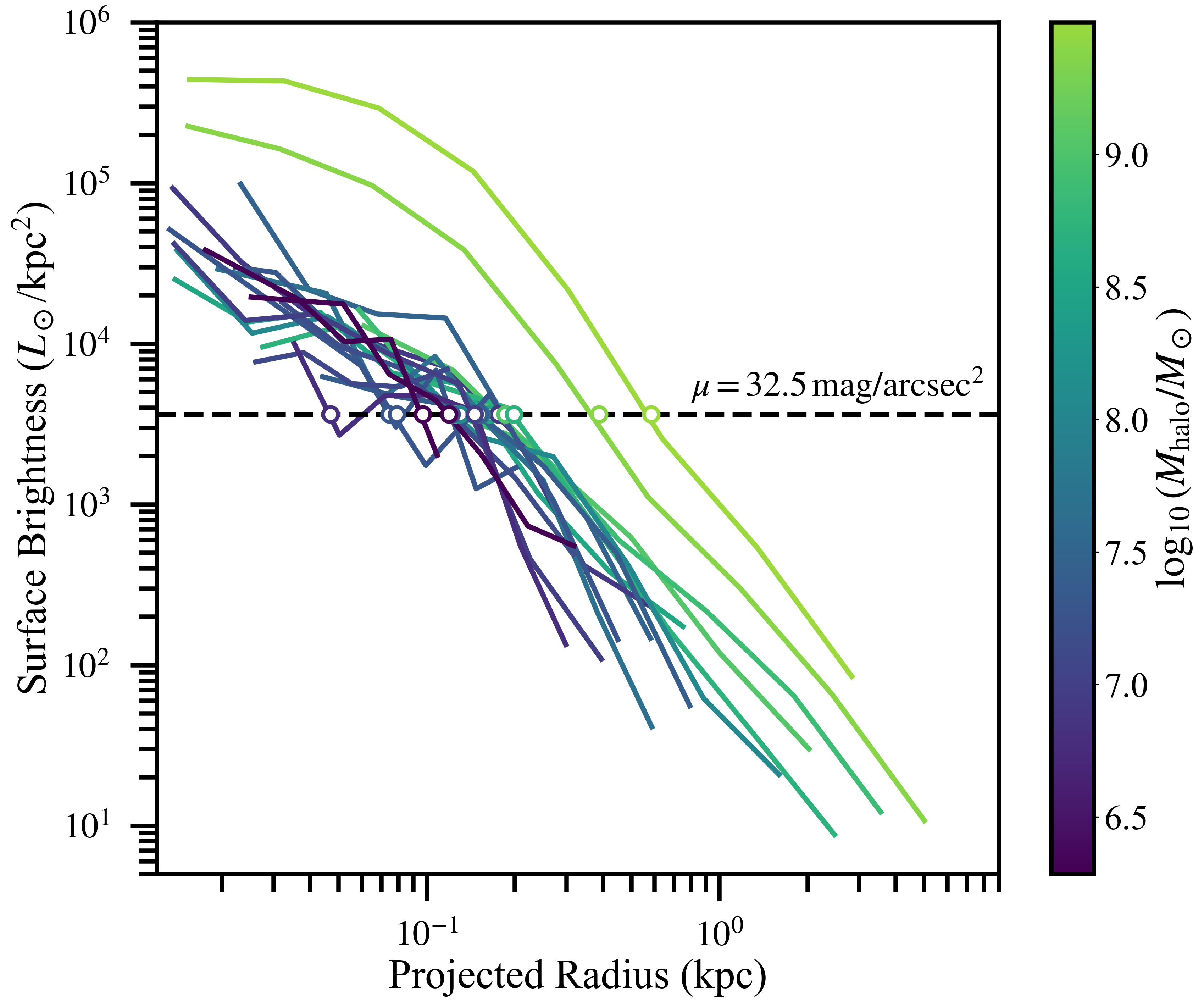} 
		\caption{Projected surface brightness profiles (assuming $M/L=1M_\odot/L_\odot$) out to $15\%$ of the halo virial radius for all ultra-faint ($\mstar <10^5~\msun$) galaxies in our sample, color-coded by dark matter halo mass. A horizontal line is shown at $32.5 \,\rm mag \, arcsec^{-2}$. We select the radius at which each profile crosses this threshold (open circles) as the new galaxy edge. In every case, this decreases the defined galaxy ``edge" of our simulated sample. We restrict our sample to all galaxies that maintain at least 10 star particles using both definitions for galaxy edge, and that have well-defined and measurable surface brightness profile above the detection threshold.
			\label{fig:profiles}
		} 
	\end{figure}
        %%%%%%%%%%%%%%%%%%%%%%%%%%%%%%%%%%%%%%%%%%%%%%%%%%%%%%%%%%%%%% 

\section{Simulations}
\label{sec:sims} 
As in \citet{Wheeler2019}, our simulated galaxy sample comes from three FIRE-2 cosmological hydrodynamic zoom-in simulations\footnote{A public release of these simulations is described in \citet{Wetzel2023} and can be found at \href{http://flathub.flatironinstitute.org/fire}{\url{http://flathub.flatironinstitute.org/fire}}} of Lagrangian volumes surrounding three central isolated low mass ($\mstar\lesssim 10^6\msun$) galaxy halos ($\mtenq$, $\mtenv$, $\mnine$) with $\mbar = 30 \msun$.\footnote{$m_{\rm dm}$ is larger by $\Omega_m/\Omega_b$} The gravitational softening for gas is adaptive, and reaches sub-pc scales. The central galaxies in these three simulations are hosted by dark matter halos with $z=0$ virial masses\footnote{We define virial overdensity with the spherical top hat approximation of \citet{Bryan1998}.} $\sim 2-10\times10^{9}\msun$.
We summarize the details of the numerical methods and initial conditions\footnote{The initial conditions (ICs) used here are publicly available at \href{http://www.tapir.caltech.edu/~phopkins/publicICs}{\url{http://www.tapir.caltech.edu/~phopkins/publicICs}}} here, and refer the reader to \citet{Hopkins2018a,Hopkins2018b} and \citet{Wheeler2019} for a more detailed accounting.

The simulations are run with the FIRE-2 implementation of star formation and stellar feedback from \citet{Hopkins2018a}, using the {\small GIZMO} \citep{Hopkins2014b}\footnote{A public version of {\small GIZMO} is available at \href{http://www.tapir.caltech.edu/~phopkins/Site/GIZMO.html}{\url{http://www.tapir.caltech.edu/~phopkins/Site/GIZMO.html}}} code. In addition to the ``meshless finite mass'' (MFM) Lagrangian finite-volume Godunov method for the hydrodynamics, FIRE-2 uses the the UV background from the December 2011 update of \citet{Faucher-Guigere2009} \footnote{Available here: \href{http://galaxies.northwestern.edu/uvb/}{\url{http://galaxies.northwestern.edu/uvb/}}}, which reionizes the universe rapidly near $z\sim 10$, completes reionization by $z=6$, and was designed to produce a reionization optical depth consistent with WMAP-7 \citep{Komatsu2011} and local sources and cooling from $T=10-10^{10}\,K$.
%The simulations use the same $\Lambda$CDM cosmology described in \citet{Wheeler2019}. 
%Details of the star formation prescription can be found in \citet{Hopkins2018a}. 
Star formation occurs in locally self-gravitating, Jeans-unstable, self-shielding/molecular gas that exceeds a critical density $n_{\rm crit}$.\footnote{$n_{\rm crit}$ is 1000 $\rm cm^{-3}$ for $\mtenq$ and $\mnine$, but was set to $10^5~\rm cm^{-3}$ for $\mtenv$} A sink-particle method is used to convert gas particles into star particles if they satisfy the above criteria. We use {\small STARBURST99} \citep{Leitherer1999} to model the stellar evolution -- each star particle has a \citet{Kroupa2002} IMF, inheriting the age and metallicity from its parent gas particle. We include stellar feedback from SNe Type Ia \&\ II, stellar mass-loss from O/B and AGB winds, photo-electric and photo-ionization heating, and radiation pressure. We follow 11 separately-tracked
elemental abundances (H, He, C, N, O, Ne, Mg, Si, S, Ca, and Fe).

\subsection{Surface Brightness Cuts and Simulated Sample}
\label{sec:sbcut} 

To identify gravitationally bound members of each halo, we run the Amiga Halo Finder \citep[AHF;][]{Knollmann2009}. For each stellar system, we compute structural and kinematic properties using two different aperture definitions: (1) $R_{\rm 15\%}$, which is a commonly used\footnote{Values from $10-20\%$ are often used in the literature \citep{Sales2010,Price2017,Hopkins2018a,Dutton2020,Rohr2022,Strawn2024}. We have further verified that there is virtually no difference between redefining this as the galaxy edge and the simple visual inspection used in \citet{Wheeler2019}.} cut including all stars within 15\% of the host halo's virial radius and (2) $R_{\rm SB}$ an isophotal cut based on the radius where the projected surface brightness profile\footnote{We do initially restrict the stellar population to be 50\% of the halo's virial radius before calculating the projected surface brightness, to eliminate extreme projection effects, i.e. stars that would be easily marked as fore- or background stars.} drops below $\mu_V = 32.5$ mag arcsec$^{-2}$ for solar absolute magnitude $\rm M_{\odot,V} = 4.83$, assuming a stellar mass-to-light ratio of $\mstar/L \approx 1\, \msun/\lsun$\footnote{Calculating $\mstar/L$ for individual UFDs using \citet{Salpeter1955} and \citet{Kroupa2002} IMF-derived stellar masses from \citet{Martin2008ApJ...684.1075M} produces a range of $\mstar/L$ values, from 1.2 - 2.6 \citep{RodriguezWimberly2019}. Additionally, measuring the IMF slopes more directly from photometric data give faint end slopes lower than both Salpeter and Kroupa IMFs ($\alpha \sim 1.25$ compared to $\alpha_S = 2.35$ and $\alpha_K = 2.3$; \citealt{Geha2013}). Given the range in derived values and the evidence for a bottom-light IMF in UFDs, we use $\mstar/L = 1$ to be consistent with current constraints. None of the trends discussed here would change with a different $\mstar/L$ ratio.} (corresponding to a physical, bolometric $0.036\,L_{\sun}\,{\rm pc}^{-2}$ for, e.g., a Plummer profile with central surface brightness $\Sigma_{\rm peak} = L/\pi\,R_{1/2}^{2}$).\footnote{This specific limit is anticipated for upcoming surveys such as the co-added Large Synoptic Survey Telescope (LSST).} Only galaxies with at least 10 star particles inside both apertures and a measurable surface brightness profile above the detection threshold are retained. We note that this SB threshold is over two orders magnitude lower than the proposed stellar surface density \citet{Trujillo2020} use as a physically motivated definition for the edge of more massive galaxies. We choose this specific threshold because we have recently begun to detect low-mass galaxies with average SB $\sim 32~\rm mag~arcsec^{-2}$ \citep{Torrealba2018,Homma2019}. Furthermore, more drastic SB cuts would have made many of our UFDs completely invisible, while this particular threshold demonstrates the effect an effective SB cut has on galaxy edges and therefor on the predictions made using this cut compared to predictions using a different cut.
%For each system, we compute the stellar half-mass radius, line-of-sight velocity dispersion, total stellar mass, mean $\rm [Fe/H]$, and enclosed true and dynamical mass using the \citet{Wolf2010} estimator. 

For each galaxy edge cut, $R_{\rm 15\%}$ and $R_{\rm SB}$, we compute stellar mass, size ($\rhalf$, the 2D-projected half-mass radius, or half-light radius), line-of-sight velocity dispersion, $\mhalf^{\rm est}$ (the estimated mass within $\rhalf$ via the \citealt{Wolf2010} mass estimator), $\mhalf^{\rm true}$ (the true total mass within the half-mass radius), and $\rm [Fe/H]$, the mean iron content of the galaxy. Fig.~\ref{fig:profiles} shows 2D projected surface brightness profiles for all ultra-faint galaxies (UFDs; $\mstar < 10^5~\msun$) in the simulations that have at least 10 star particles within the aperture region, color-coded $\mhalo$, the mass (in $\msun$) of the dark matter halo. We are not analyzing the two most massive centrals, $\mtenq$ and $\mtenv$, that are the dominant galaxies in the $\sim 10^{10}~M\odot$ halos. This is because these two classical dwarf spheroidals (dSphs) have most of their stellar population above the SB cut, and so $R_{\rm SB} > R_{15\%}$. We refer the reader to \citet{Wheeler2019} for details on the two most massive dSphs. There are also two ultra-faints far more massive than the others, $\mnine$ and another isolated ultra-faint in the $\mtenv$ Lagrangian region. Of the 20 UFDs that satisfy the 10 star particle requirement, 10 are satellites of the two dSPh hosts ($\mtenq$ and $\mtenv$), two are satellites of other UFDs, and 8 are isolated. We remove one satellite of $\mtenq$ due to excessive contamination from stars in the stellar halo of the more massive dSph, leaving 19 ultra-faint galaxies in the sample.

 %        %%%%%%%%%%%%%%%%%%%%%%%%%%%%%%%%%%%%%%%%%%%%%%%%%%%%%%%%%%%%%% 
	% \begin{figure}
	% 	\includegraphics[width=0.48\textwidth]{SMHM_15_vs_SB_Clean.png} 
	% 	\caption{Stellar mass-halo mass ($M_\star-M_{\rm halo}$) relation for all central/satellite (star/triangle) galaxies defining the galaxy edge using a fiducial cut at $15\%$ of $\rvir$ ($R_{15\%}$; open symbols) and using a cut at an effective surface brightness limit of $\mu \sim 32.5$ mag arcsec$^{-2}$ ($R_{\rm SB}$; closed symbols). We compare to the extrapolated abundance matching (AM) relations from\citet{Garrison-Kimmel2014a} and \citet{Brook2014} with 0.7-dex scatter, as well as extrapolations from the AM relations of \citet{Moster2013} and \citet{Behroozi:2013} without the scatter. The points are color-coded by the log displacement between the resulting stellar mass with each method. Star symbols show results for isolated galaxies and triangles show satellites of other low mass galaxies in the simulation (in almost every case they are satellites of the most massive central). Due to our requirement that our galaxies have at least 10 star particles, we see an artificially imposed flattening of the relation at low halo mass. While most of our simulated galaxies are well below detection limits for the observations used to calibrate the relations, we do note that the lower masses due to the surface brightnesses cut moves our galaxies into slightly better agreement with these extrapolations.
	% 		\label{fig:SMHM}
	% 	} 
	% \end{figure}
 %        %%%%%%%%%%%%%%%%%%%%%%%%%%%%%%%%%%%%%%%%%%%%%%%%%%%%%%%%%%%%%% 

\section{Results}
	\label{sec:results}

             %%%%%%%%%%%%%%%%%%%%%%%%%%%%%%%%%%%%%%%%%%%%%%%%%%%%%%%%%%%%%% 
	\begin{figure}
		\includegraphics[width=0.48\textwidth]{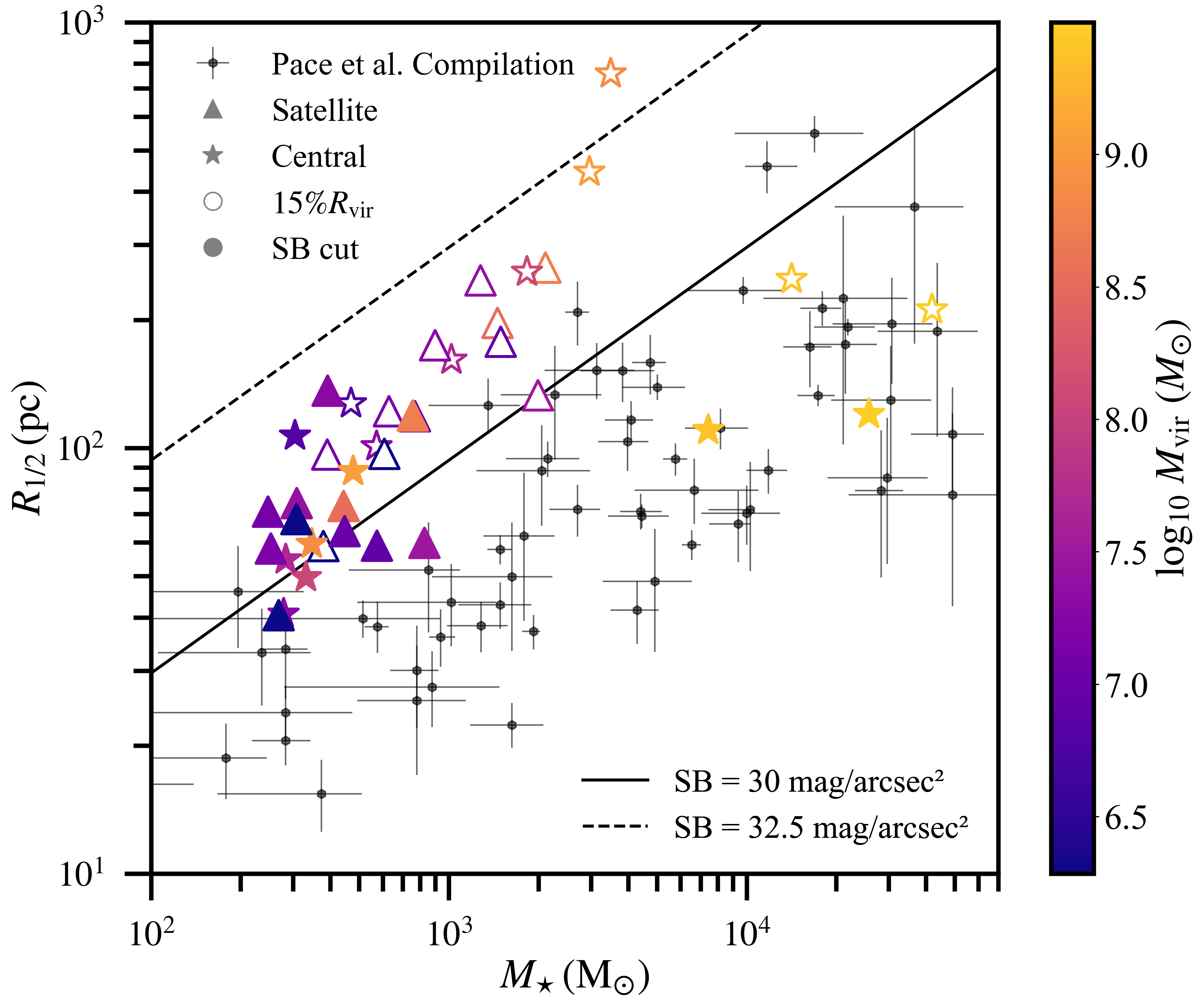} 
		\caption{2D (projected) half-stellar-mass radii ($\rhalf$) versus $\mstar$ for all simulated galaxies in the sample. Open symbols show values calculated assuming the galaxy edge is $R_{15\%}$, while solid symbols use the $R_{\rm SB}$ cut described in Section \ref{sec:sbcut}. Marker shapes indicate centrals (stars) and satellites (triangles). Observed MW UFDs are compiled from \citet{Pace2024}. The solid line is a typical surface brightness limit of these surveys, $30 \rm \, mag \,arcsec^{-2}$ for $M/L\approx 1~M_\odot/L_\odot$. The dashed line shows a surface brightness limit of $32.5 \rm \, mag \, arcsec^{-2}$ more comparable to future surveys such as LSST. Galaxies are color-coded by their $V_{300}$ values. 
		\label{fig:masssize}
        }
	\end{figure}
        %%%%%%%%%%%%%%%%%%%%%%%%%%%%%%%%%%%%%%%%%%%%%%%%%%%%%%%%%%%%%% 

% Fig. \ref{fig:SMHM} shows the location of the galaxies on the stellar-mass halo-mass (SMHM) relation for the $R_{\rm 15\%}$ and $R_{\rm SB}$ cuts. 
The galaxy edge defined by the surface brightness cut ($R_{\rm SB}$; open circles in Fig.~\ref{fig:profiles}) decreases the stellar mass of each galaxy, while leaving the halo mass unchanged. Our simulated galaxies occupy dark matter halos with $2\times 10^6 \lesssim \mhalo/\msun \lesssim 3\times 10^9$. The stellar masses range from $380$-$1.4\times 10^4~\msun$ using our $R_{15\%}$ cut, and from $250$-$2.6\times 10^4~\msun$ using our $R_{\rm SB}$ cut.
While none of our simulated galaxies are satellites of anything more massive than the two most massive galaxies in the figure ($\mhalo \lesssim 10^{10}\msun$), we represent centrals with star symbols and satellites with triangles in the figures that follow. 
%Lines show constraints from abundance matching prescription extrapolations from higher mass. Noting that $M_{z=0}^{\rm halo}$ is typically about a factor of 2 smaller than $\mpeak$, we see that imposing the artificial limit of 10 star particles creates an apparent flattening of the relation \citep{Sawala2015}. Dropping this limit and looking at the SMHM relation for any halo that forms at least one single star particle does alleviate the flattening, but introduces a new problem of single star particle ``galaxies" formed in dark matter halos below the atomic cooling limit (Wheeler et al.~\textit{in prep}). 
%Each point is color-coded by the log of the stellar mass change between this and the $R_{\rm 15\%}$ cut (open symbols). For galaxies of comparable $\mstar$, there is a slight tendency for a larger change in stellar mass for galaxies with larger $\mhalo$, due to the flatter relationship between $M_\star^{\rm SB}$ and $\mvir$ when compared to the slightly steeper $M_\star^{\rm 15\%}$-$\mhalo$ trend. 

%%%%%%%%%%%%%%%%%%%%%%%%%%%%%%%%%%%%%%%%%%%%%%%%%%%%%%%%%%%%%%
\begin{figure*}
    \centering
    \includegraphics[width=0.48\textwidth]{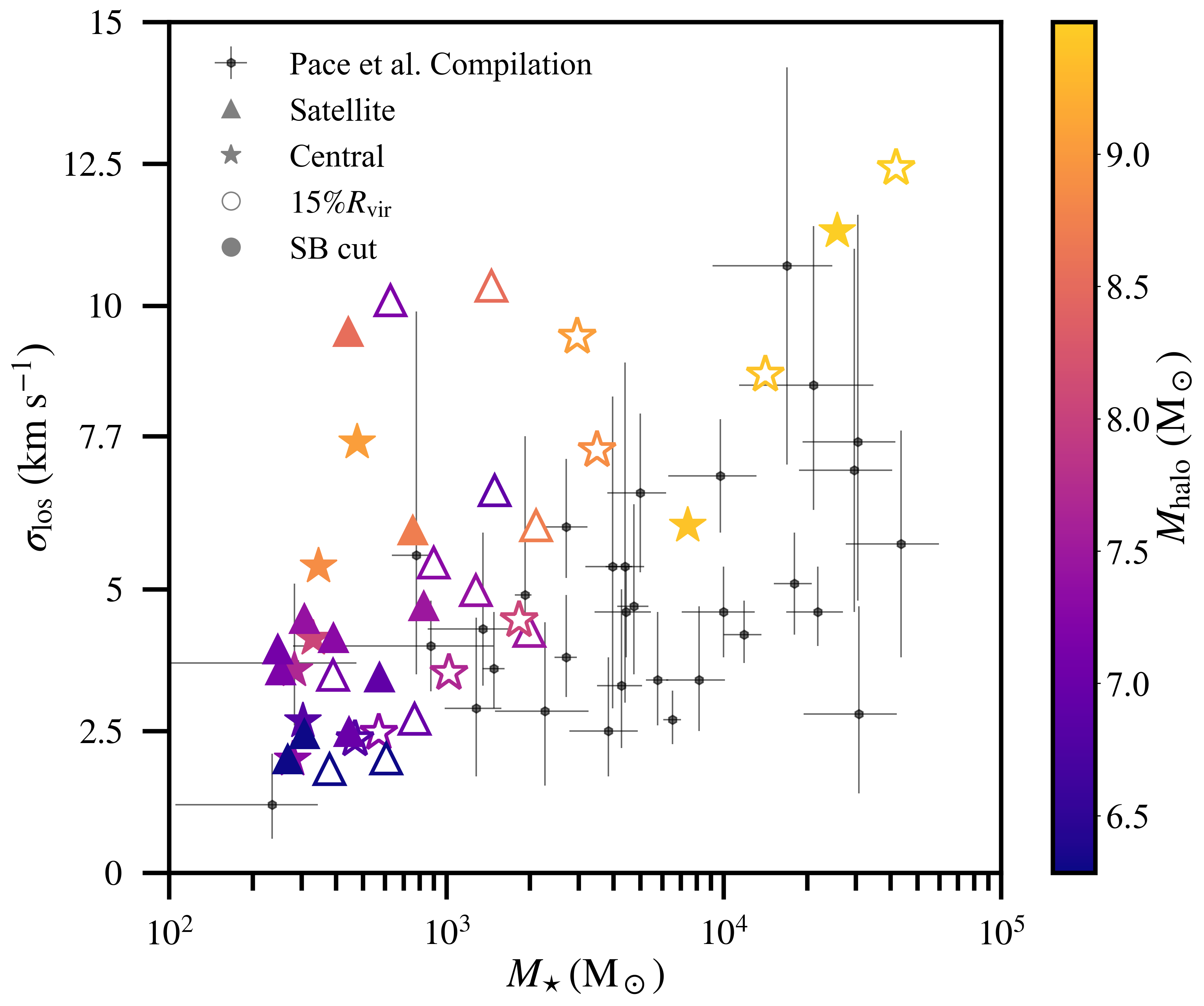}
    \includegraphics[width=0.48\textwidth]{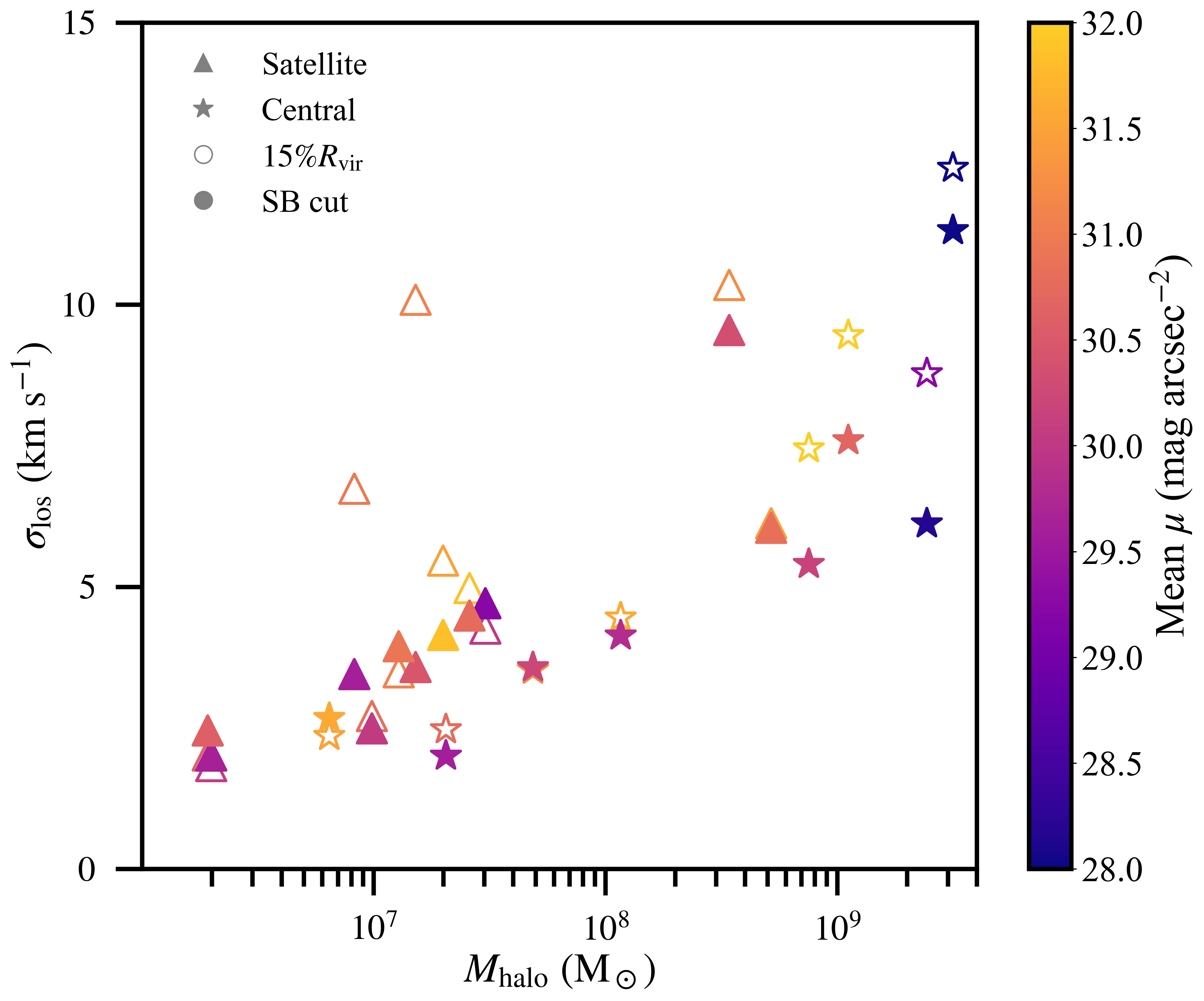}
    \caption{
    \textbf{Left:} Line of sight velocity dispersion versus stellar mass within the galaxy ``edge" for the $R_{\rm 15\%}$ (open symbols) and $R_{\rm SB}$ cut (solid symbols), color-coded by dark matter halo mass. Marker shapes indicate centrals (stars) and satellites (triangles). Data are from a compilation by \citet{Pace2024}, and are limited to $\mstar < 10^5~\msun$. The $R_{\rm SB}$ cut moves the predicted $\sigma_{\rm los}$ towards lower values, and closer to observations. 
    \textbf{Right:} Line of sight velocity dispersion versus dark matter halo mass, color-coded by surface brightness. Using the $R_{\rm SB}$ cut, a weak trend exists between halo mass and surface brightness for galaxies with similar $\sigma_{\rm los}$, such that lower mass halos are more difficult to see. However the opposite trend exists for the $R_{\rm 15\%}$ cut.
    }
    \label{fig:vdisp}
\end{figure*}
%%%%%%%%%%%%%%%%%%%%%%%%%%%%%%%%%%%%%%%%
\subsection{Galaxy Sizes}

Fig.~\ref{fig:masssize} shows the mass-size relationship for ultra-faint galaxies ($\mstar < 10^5~\msun$) in the Local Volume (LV) compiled by \citet{Pace2024}.\footnote{\url{https://github.com/apace7/local_volume_database}} The Figure also shows overall effective surface brightness detection limits of $\rm \mu_V = 30~mag~arcsec^{-2}$ (solid line), corresponding to the surface brightness detection limit for the Sloan Digital Sky Survey \citep[SDSS; ][]{Willman2002}, and the stricter limit of $\rm 32.5~mag~arcsec^{-2}$ (dashed line), more comparable to future
surveys such as LSST \citep{LSST2024}. The fact that galaxies follow a tight relationship between their stellar mass and their size as determined by a half-light radius has been known for some time \citep{Shen2003,Kravtsov2013,Mercado2025}, and has been shown to extend to $z\sim 3$ \citep{Miller2019,Mowla2019,Nedkova2021}. However, the tendency of the data to track lines of constant surface brightness, combined with recent detections of ultra-diffuse galaxies and other extremely low-surface brightness objects such as Antlia 2 \citep{vanDokkum2015,Torrealba2018}, hints at a potentially hidden population of galaxies with extended features, and surface brightness lower than current detection limits \citep{Bullock2010}.

Alongside the observed data\footnote{We note that \citet{Pace2024} derive their $\mstar$ values using $\mstar/L=2~\rm M_\odot/L_\odot$. Re-deriving their results with a lower $\mstar/L$ ratio would shift their values to the right slightly, leading to less overlap with our results, but would not change the overall trends described here.}, we plot predictions from our simulations for the mass-size relation using the $R_{\rm 15\%}$ (open symbols) and $R_{\rm SB}$ cuts (closed symbols), color-coded by $\mhalo$. We use the galaxy edge defined by the surface brightness cut to re-calculate the half-mass radius for each galaxy, as well as the stellar mass. The shift to lower $\mstar$ and particularly lower $\rhalf$ between the two methods is striking. Adopting the $R_{\rm SB}$ cut typically reduces galaxy sizes by over $50\%$, with substantial halo-to-halo variation ($\pm 20\%$ relative to the original size). The two most massive UFDs, $\mnine$ and $\mtenvB$, already overlap with the data using the $R_{\rm 15\%}$ cut, but move significantly closer to the to the mean observational relation after applying the $R_{\rm SB}$ cut. Furthermore, the lowest mass galaxies from \citet{Wheeler2019} occupying the region in the Figure near the $\rm 32.5~mag~arcsec^{-2}$ line undergo a dramatic shift in their location on the figure due to the large decrease in their half-mass radii--even overlapping with the data at the lowest stellar masses. While none of the open symbols reach $\lesssim 30\rm ~pc$, most are well below $100\rm ~pc$, compared to nearly all of the points having $\rhalf \gtrsim ~100\rm ~pc$ using the $R_{\rm 15\%}$ cut.

    While \citet{Wheeler2019} did not predict a tight correlation between $\rhalf$ and $\mstar$ for the UFDs sampled in that paper (galaxies that formed $>100$ star particles), a steep mass-size relation can be seen using the $R_{\rm 15\%}$ cut (open symbols) for the lowest-mass galaxies in the sample (more than 10 star particles, but with $\mstar \lesssim 10^4\msun$). This relationship is a direct result of using a fixed fraction of $\rvir$ to define the galaxy edge, combined with the fact that $\mhalo \propto \rvir^3$, and finally factoring in the stellar mass-halo mass relation (despite increased scatter at the lowest masses \citealt{Munshi2021}).
 The relation disappears, becoming a scatter plot at low $\mstar$ and $\rhalf$, when the galaxy edge cut based on surface brightness, $R_{\rm SB}$, is imposed (solid symbols). 
The increase in overall surface brightness due to the $R_{\rm 156\%}$ cut, and particularly the improved overlap between the observed data and the simulated points with the $R_{\rm SB}$ cut, supports the idea that observations may only be sensitive to the ``bright'' core of more massive, extended, objects and that observed UFDs all have diffuse (and relatively massive) outer halos that are currently invisible. Low-mass galaxies are predicted to be \textit{more} extended near a massive host \citep{Mercado2025}, further strengthening the idea that we are likely missing much of the observed structure at large radii, or it has been tidally stripped.

\subsection{Velocity Dispersion}
\label{sec:vdisp}
	
We plot line of sight velocity-dispersion, $\sigma_{\rm los}$ calculated along the z-axis, for all stars within the galaxy ``edge" for the $R_{15\%}$ (open symbols) and $R_{\rm SB}$ cuts (closed symbols) versus stellar mass in the left panel of Fig. \ref{fig:vdisp}, alongside observed UFD values from \citet{Pace2024}. The simulated points are color-coded by $\mhalo$. Imposing the surface brightness cut at $R_{\rm SB}$ eliminates stars at large galactocentric distances, which causes the galaxies to shift to the left, due to their decreased stellar mass. We then calculate the velocity dispersion of a smaller, more centrally concentrated stellar population. All but some of the lowest mass galaxies shift towards lower velocity dispersions, although most move largely horizontally, due to the flat velocity dispersion profiles (although several satellites move significantly lower). Using the more extended $R_{\rm 15\%}$ cut yields a mean value of $\langle\sigma^{\rm 15\%}_{\rm los}\rangle = 6.36\rm ~km~s^{-1}$, with 10/19 of the values below $5~\rm km~s^{-1}$. The $R_{\rm SB}$ cut gives $\langle\sigma^{\rm SB}_{\rm los}\rangle = 5.36\rm ~km~s^{-1}$, with three more now below $5~\kms$. The most massive dark matter halos in our simulation are just under $10^{10}~\msun$, and even many of our centrals have $\sigma_{\rm los}<5~\kms$, confirming that these galaxies are capable of having kinematically cold stellar populations without the presence of a more massive host \citep{Penarrubia2008}, despite their extended sizes \citep{Mutlu-Pakdil2019}.

            %%%%%%%%%%%%%%%%%%%%%%%%%%%%%%%%%%%%%%%%%%%%%%%%%%%%%%%%%%%%%% 
	\begin{figure}
		\includegraphics[width=0.48\textwidth]{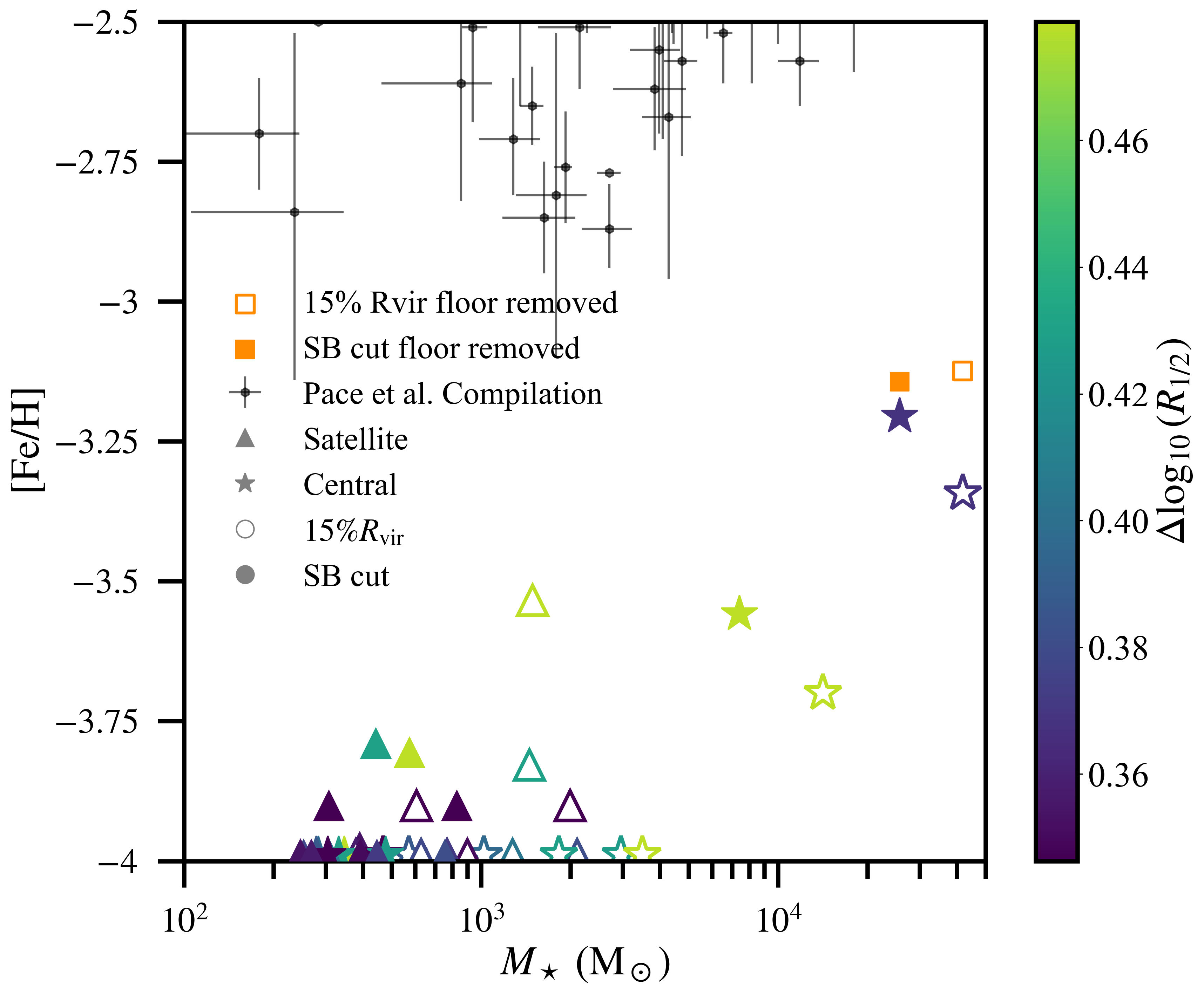} 
		\caption{ Stellar mass ($\mstar$) versus metallicity ([Fe/H]) our simulated galaxies using a galaxy edge defined as $15\% ~\rvir$ ($R_{15\%}$; open symbols) and using a surface brightness cut at $\mu = 32.5\rm mag~arcsec^{-2}$ ($R_{\rm SB}$; closed symbols). As in \citet{Wheeler2019}, the UFDs at $\rm [Fe/H] \sim -4$ are at the baseline initial metallicity of the simulation. The two most massive UFDs show slightly improved agreement with the updated data from \citet{Pace2024}, shown as points with error bars for both methods of determining galaxy edge, but the surface brightness cut moves the simulated points even closer.
			\label{fig:FeH}
		} 
	\end{figure}
        %%%%%%%%%%%%%%%%%%%%%%%%%%%%%%%%%%%%%%%%%%%%%%%%%%%%%%%%%%%%%%	

%%%%%%%%%%%%%%%%%%%%%%%%%%%%%%%%%%%%%%%%%%%%%%%%%%%%%%%%%%%%%%
\begin{figure*}
    \centering
    \includegraphics[width=0.48\textwidth]{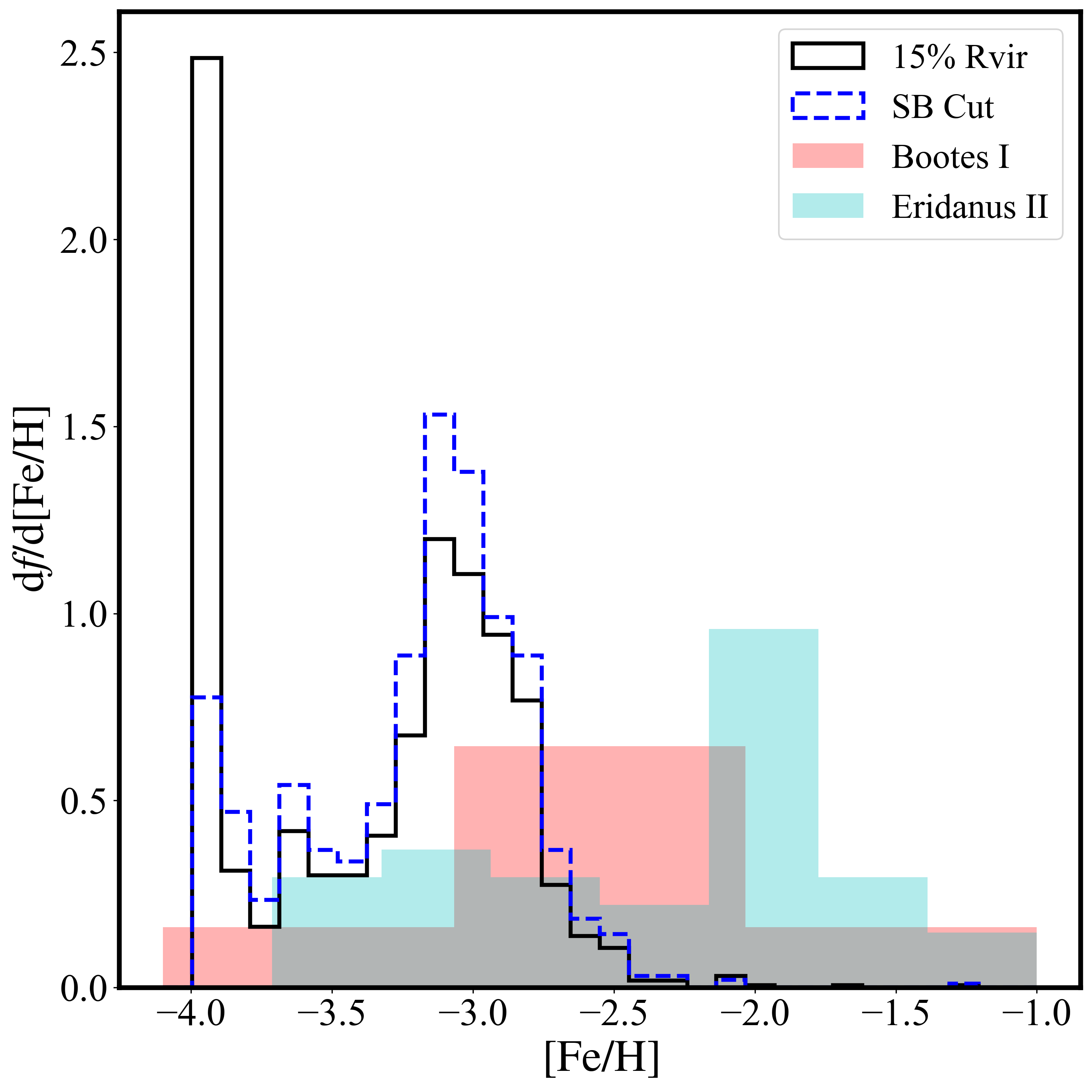}
    \includegraphics[width=0.48\textwidth]{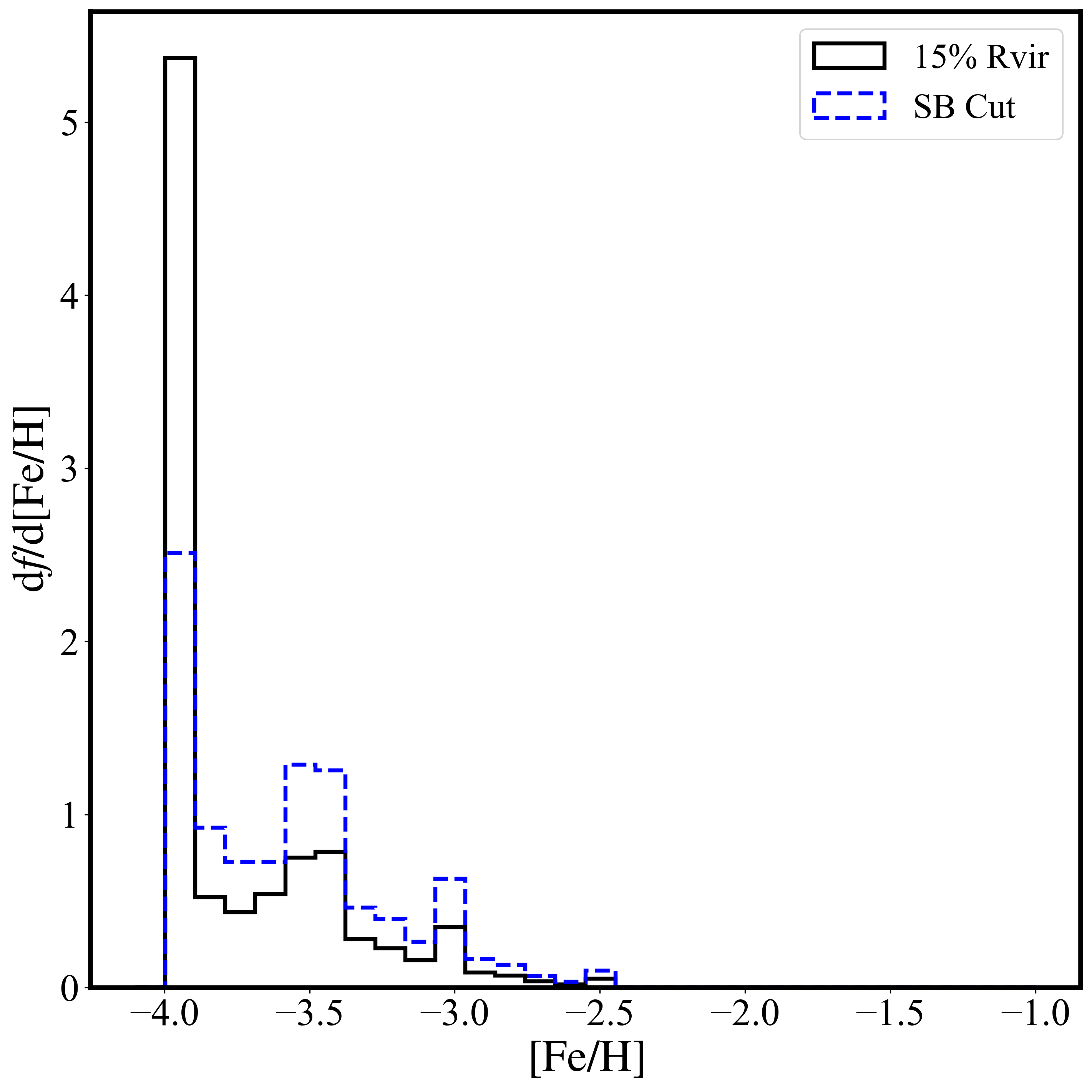}
    \caption{
    \textbf{Left:} Metallicity distribution function (MDF) for our most massive ultra-faint galaxy, $\mtenvB$, for the $R_{15\%}$ (solid black line) and surface brightness limit ($R_{\rm SB}$; dashed blue line). MDFs for Bootes 1 (Eridanus 2) are also shown as red (cyan) shaded histograms (Rodriguez Wimberly \textit{in prep}). Many of the stars remain at the metallicity floor of the simulation, and those stars are preferentially removed from the outer edge of the galaxy when performing the $R_{\rm SB}$ cut, shifting the mean [Fe/H] upwards.
    \textbf{Right:} Same but for the second most massive galaxy, $\mnine$.
    }
    \label{fig:MDF}
\end{figure*}
%%%%%%%%%%%%%%%%%%%%%%%%%%%%%%%%%%%%%%%%%%%%%%%%%%%%%%%%%%%%%%

If the stars in UFDs act as massless tracers of the dark matter halo's potential, then the stars in a galaxy that forms in a low-concentration or low-mass halo must spread out to maintain the same velocity dispersion. Galaxies with similar stellar velocity dispersions could have very different effective radii, depending on the depth of their host dark matter potential \citep{Bullock2010}. Galaxies with larger radii for a given stellar mass would then have lower effective surface brightnesses, which could cause these systems to go undetected. So-called ``stealth'' galaxies would preferentially populate the low-concentration, low-mass tail of the halo distribution at fixed stellar velocity dispersion or luminosity, and may evade detection in current surveys \citep{Bullock2010}. Fig. \ref{fig:vdisp} shows that, for our simulated galaxies, $\sigma_{\rm los}$ tends to rise with stellar mass (left panel) and halo mass (right panel), while the observed galaxies have a flatter trend of $\sigma_{\rm los}$ at least with $\mstar$. This makes comparing surface brightness at fixed velocity dispersion difficult. It is difficult to see any strong trend between halo mass and surface brightness, but if we look at centrals only with $\sigma_{\rm los}<7.5~\kms$, using the $R_{\rm SB}$ cut, we find a best-fit relation of the form
$\mu = -0.86\rm ~log_{10}(\mhalo)+36.84$, 
with a coefficient of determination
$R^2 = 0.575$, confirming a weak anti-correlation between halo mass and surface brightness: galaxies of similar velocity dispersion in lower massive halos tend to exhibit lower mean surface brightnesses. However, performing the same calculation on the values derived from the $R_{\rm 15\%}$ cut, we see the opposite trend, $\mu = 0.93\rm ~log_{10}(\mhalo)+24.46$, 
with a coefficient of determination
$R^2 = 0.617$. Unfortunately the extremely low sample size makes any conclusions difficult to draw. 

\subsection{Mass metallicity Relation}
\label{MZR}

The lowest mass galaxies that form in the extreme high resolution limit of cosmological hydrodynamic simulations are deficient in metals compared to ultra-faint Milky Way satellites in the same mass range \citep{Maccio2017, Wheeler2019, Agertz2019, Ko2024}. A possible cause of the break between simulations and observations at the low end of the mass-metallicity-relation (MZR) could be that feedback ejects too many metals from the simulated galaxies, and that the few bursts of star formation that do occur do so from un-enriched gas. Another potential cause of the discrepancy could be the lack of a more massive Milky Way in the highest resolution cosmological simulations of ultra-faints. Simulations of a $\mbar = 880\msun$ Milky Way run with FIRE-2 physics to $z=5$ show that, although many UFDs remain at the metallicity floor, some do become enriched to simulated levels due to the galaxies forming in regions with higher star formation rates, and therefor increased metal production \citep{Wheeler2019}. Observations of galaxies with $\mstar = 10^{6.1}-10^{9.3}\msun$ in low density environments have lower metallicity than Local Group (LG) galaxies in the same mass range \citep{Heesters2023,Muller2025}. Another possible solution to the discrepancy could be updating Type Ia supernova rates to be metallicity-dependent, such that low-mass galaxies have higher specific SN Ia rates (rate per unit stellar mass) compared to Milky Way-mass galaxies \citep{Brown2019}. \citet{Gandhi2022} run FIRE-2 simulations with metallicity-dependent SN Ia rates, finding that they significantly improve the agreement between the simulated galactic stellar masses and observations. \citet{Prgomet2022} find that incorporating a top-heavy IMF at low metallicity in simulations increases the efficiency of feedback in regulating stellar mass, while retaining sufficient metals due to increased metal production, aligning better with observed MZR trends. Most likely, the solution contains some form of inclusion of population III star models, which have been shown to raise the metallicity of low-mass galaxies \citep[][Gandhi et al. \textit{in prep}]{Jeon2021,Prgomet2022,Sanati2023}.

We test whether or not eliminating outer stars below the effective surface brightness cut from the UFDs can alleviate this problem. Low-mass galaxies exhibit inverted age and metallicity gradients \citep{delPino2015,Graus2019,Belland2020,Chiti2021,Mercado2021}, with young stellar populations forming primarily \textit{in-situ} in the center and older populations either gradually pushed outward over time due to the fluctuations of the potential well created by bursty star formation \citep{ElBadry2017}, or deposited by mergers with less massive, ancient systems \citep{Benitez-Llambay2015,Graus2019,Rey2019,Goater2024,Ko2024}. In either case, removing these outer stars could conceivably raise the age and/or mean iron content of the galaxies. 

%%%%%%%%%%%%%%%%%%%%%%%%%%%%%%%%%%%%%%%%%%%%%%%%%%%%%%%%%%%%%%
\begin{figure*}
    \centering
    \includegraphics[width=0.48\textwidth]{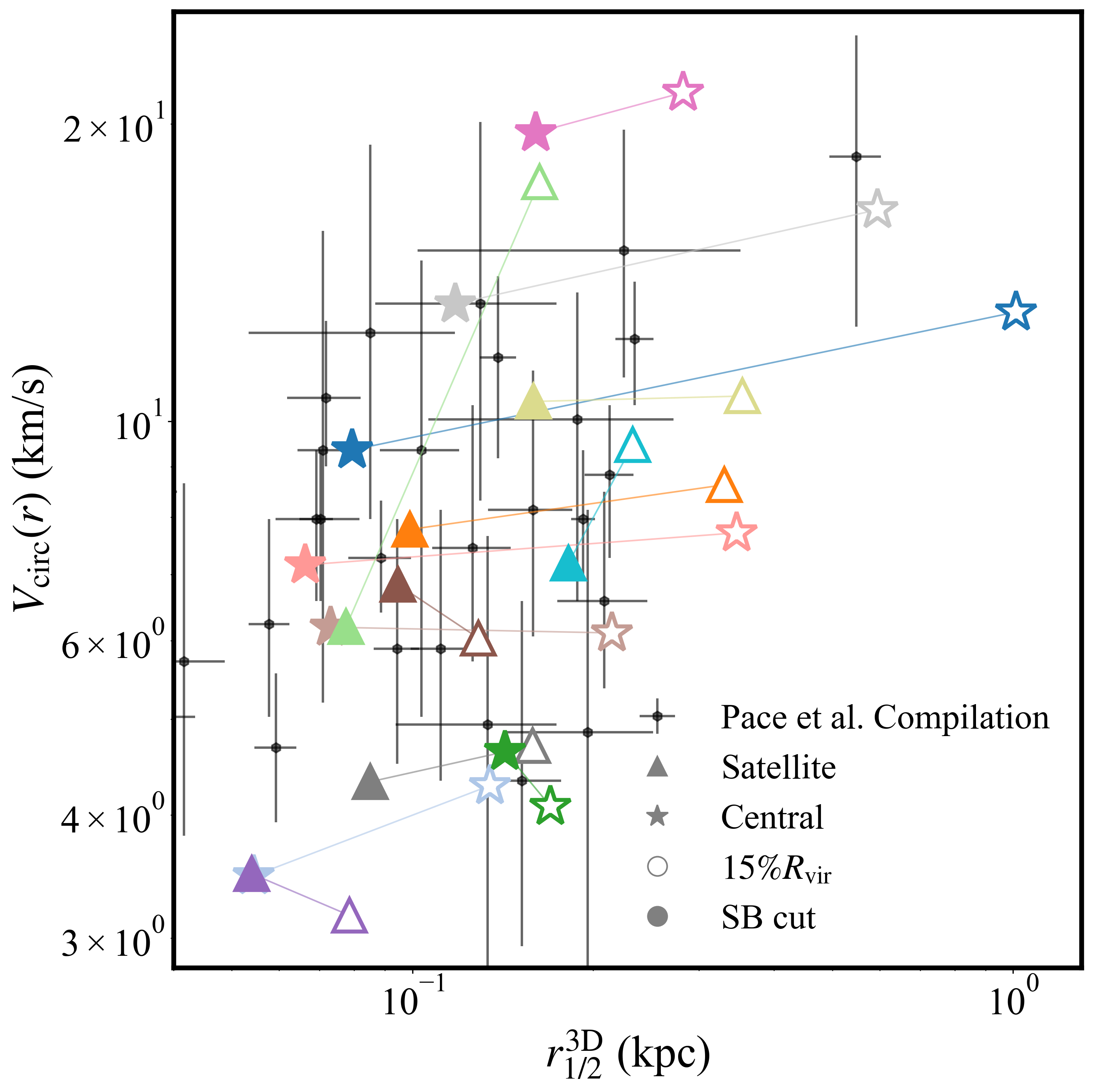}
    \includegraphics[width=0.48\textwidth]{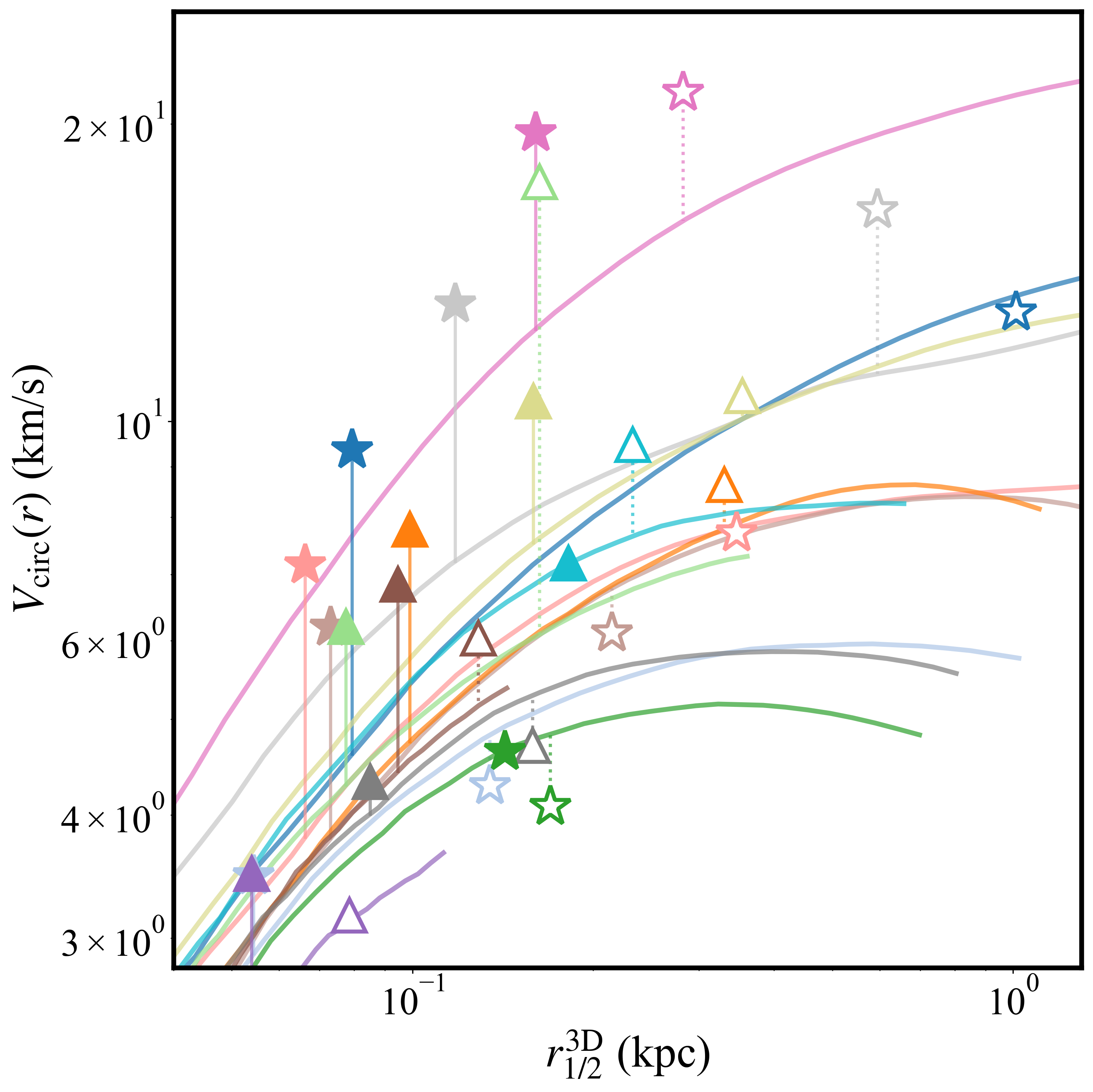}
    \caption{
    \textbf{Left:} \citet{Wolf2010} mass estimator evaluated at the 3D de-projected half-light radius using the $R_{15\%}$ cut (open symbols) and the $R_{\rm SB}$ cut (closed symbols) for the same galaxy. Marker shapes indicate centrals (stars) and satellites (triangles). Thin lines connect matched pairs for each halo. Data from \citet{Pace2024} is shown as points with error bars.
    \textbf{Right:} Circular velocity profiles for galaxies in the m10v simulation. Solid lines show the total circular velocity profile computed from the total dark matter, gas, and stellar mass enclosed within the 3D radius.  Color-matched markers denote the \citet{Wolf2010} mass estimator as in the left panel.
    }
    \label{fig:TBTF}
\end{figure*}
%%%%%%%%%%%%%%%%%%%%%%%%%%%%%%%%%%%%%%%%

We investigate this in Fig. \ref{fig:FeH}, where we have plotted the mean $\rm [Fe/H]$ for each galaxy versus its stellar mass for both galaxy edge models. We plot the \citet{Pace2024} compilation of observations as data points with error bars, and demonstrate the well-known discrepancy referenced above, although the \citet{Pace2024} compilation includes points from \citet{Simon2019}, which are slightly lower in [Fe/H] than the original \citet{Kirby2013} and \citet{Vargas2014} datasets used in \citet{Wheeler2019}. However, as in \citet{Wheeler2019}, most of the lowest-mass UFDs lie at the metallicity floor of the simulation ($\rm [Fe/H] \sim -4$), failing to enrich enough stars beyond the initial conditions. This is true even when performing more restrictive $R_{\rm SB}$ cut (solid symbols), likely due to the uniformly ancient stellar populations and lack of discernible metallicity gradients in these galaxies.

In the two most massive UFDs, defining the galaxy edge with the more restrictive $R_{\rm SB}$ cut does produce a slight shift towards higher metallicity, but in every case under-predict [Fe/H] by over an order of magnitude for galaxies in the same stellar mass range.
%, 
%suggesting that this may be \textit{a part} of the overall solution to this issue. 
We show why this occurs in the metallicity distribution functions for our two most massive galaxies, $\mtenvB$ (left panel) and $\mnine$ (right panel), in Fig. \ref{fig:MDF}, plotted for the $R_{15\%}$ (solid) and $R_{\rm SB}$ (dashed) cuts.
In both models, a non-negligible fraction of stars remain at the metallicity floor, having formed out of unenriched gas.\footnote{$\mtenv$ was run with a metallicity floor an order of magnitude lower than the other two simulations. To make the metallicity floors consistent between simulations, we shifted the [Fe/H] values in $\mtenv$ by adding just enough iron to match the minimum abundance of the higher floor. This adjustment preserves the physical meaning of [Fe/H] while allowing for a fair comparison across simulations with different enrichment limits.}  Some of this may be due to later bursts forming from pristine gas, having blown out the enriched gas with SNe explosions, but most of the star formation in these two galaxies occurs in small halos within the first few hundred $\myr$ of the simulation, so the more likely issue is that those first generation, unenriched stars stick around, while true Pop III stars would have exploded. Removing stars at the outer edge of the galaxy that fall outside $R_{\rm SB}$ preferentially removes stars at the extreme low-metallicity tail of the simulation, shifting the average to higher [Fe/H]. This suggests that missing extended stellar halos may bias observed [Fe/H] measurements higher than their true values, but a surface brightness cut can only be \textit{part} of the solution to the MZR discrepancy. Removing all stars with $\rm [Fe/H] <-3.9$ from $\mtenvB$ shifts the mean [Fe/H] for \textit{both models} up to $\rm [Fe/H]\sim-3$ (open and closed square, Fig. \ref{fig:FeH}), suggesting that models for pop III star formation may also have an impact. In the left panel, we also plot the MDF for Bootes I (magenta shaded; $\mstar = 10^{4.64}~\msun$) from Rodriguez-Wimberly \textit{in prep}, and Eridanus II (cyan shaded; $\mstar = 10^{5.08}~\msun$) from \citet{Fu2023}. Even neglecting the $\rm [Fe/H] = -4$ stars, the MDFs of the two observed UFDs only partially overlap the MDF of $\mtenv$. Observations of 13 Milky Way UFDs from \citet{Fu2023} show that while the MDFs of observed ultra-faints don't have such large $\rm [Fe/H] \lesssim -4$ populations, they do have many stars at this extremely low metallicity, particularly when allowing for stars with lower signal-to-noise. This suggests that observations are potentially missing a larger fraction of these extremely low metallicity stars, and improved observations will continue to push the data points towards lower [Fe/H].

\subsection{Dark Matter Halo Mass Estimates}

Determining the dark-matter content of low-mass galaxies is a complex and important task, with implications for the missing satellites problem \citep{Klypin1999,Moore1999}, the core-cusp controversy \citep{Flores1994,Moore1994}, and other open questions such as ``Too Big to Fail" \citep{Boylan-Kolchin2011}, and the dark matter deficient galaxies problem \citep{vanDokkum2019,Moreno2022}. A common method for estimating the dynamical mass within the visible extent of ultra-faint galaxies is the \citet{Wolf2010} mass estimator,

        \begin{equation}
        \left(\frac{M^{\rm est}_{1/2\\}}{M_\odot}\right)\approx 930 \left(\frac{\sigma_{\rm los}}{\rm km~s^{-1}}\right)^2\left(\frac{(4/3)~\rhalf}{\rm pc}\right),
        \label{eq:wolf}
        \end{equation}

%%%%%%%%%%%%%%%%%%%%%%%%%%%%%%%%%%%%%%%%%%%%%%%%%%%%%%%%%%%%%%
% \begin{figure*}
%     \includegraphics[width=0.99\textwidth]{vcirc_wolf_m10v_lines_10parts_Clean_DATA_long.png}
%     \caption{
%     \textbf Circular velocity profiles for galaxies in the m10v simulation. Solid lines show the total circular velocity profile computed from the total dark matter, gas, and stellar mass enclosed within the 3D radius. Color-matched markers denote the \citet{Wolf2010} mass estimator evaluated at the 3D de-projected half-light radius using the $R_{15\%}$ cut (open symbols) and the $R_{\rm SB}$ cut (closed symbols) for the same galaxy. Marker shapes indicate centrals (stars) and satellites (triangles). Thin lines connect matched pairs for each halo. Data from \citet{Pace2024} shows values for all galaxies with $\mstar \lesssim 10^5~\msun$. Removing the stellar halo ($R_{\rm SB}$ cut), moves mass estimates closer to observations, but farther from the true enclosed mass.
%     }
%     \label{fig:TBTF}
% \end{figure*}
%%%%%%%%%%%%%%%%%%%%%%%%%%%%%%%%%%%%%%%%%%%%%%%%%%%%%%%%%%%%%%

   \noindent where the mass estimated is $\mhalf^{\rm est}$, the mass within then de-projected half-light radius, $4/3~\rhalf$, and $\sigma_{\rm los}$ is the line of sight velocity dispersion for all stars within the galaxy. Both $\rhalf$ and $\sigma_{\rm los}$ are dependent on whether or not the stellar halo is detected. If the half-mass radii of the ultra-faints around the MW are drastically under-predicted due to missing stars at large radii, this could have a significant impact on the mass estimates. Fig. \ref{fig:TBTF} shows circular velocity profiles for galaxies in $\mtenv$\footnote{We choose this subsample as it is the largest selection of UFDs, 14 out of our 19, and half of them are centrals. Adding more galaxies would clutter the Figure, and show the same conclusion.}, calculated using the total enclosed mass, as a function of the 3D distance from the galaxy center. Each circular velocity profile is arbitrarily color-coded
    to match $\rhalf$ estimates from \citet{Wolf2010} for the same simulated galaxy, shown for the $R_{\rm 15\%}$ (open symbols) and $R_{\rm SB}$ (closed symbols) cuts. For many galaxies, changing $\rhalf$ does not significantly change $\vcirc = \sqrt3\sigma_{\rm los}$, a result of the flatness of the velocity dispersion profiles. However, the de-projected half-mass radii \textit{do} change. In most cases this pulls the estimate left of the true value, such that, by missing the stellar halo of these ultra-faint galaxies, we would assume they occupy much more massive, or at least denser, dark matter halos than they actually do. Data taken from the \citet{Pace2024} compilation for any galaxy with a stellar mass less than $\mstar \lesssim 10^5~\msun$ demonstrate a greater variety in both $\rhalf$ and $\vcirc$ than do our simulations, with the most notable discrepancies at small half-light radii, where some galaxies achieve circular velocities as high as $10~\kms$ within just $20\rm ~pc$.

           %%%%%%%%%%%%%%%%%%%%%%%%%%%%%%%%%%%%%%%%%%%%%%%%%%%%%%%%%%%%%% 
	\begin{figure}
		\includegraphics[width=0.48\textwidth]{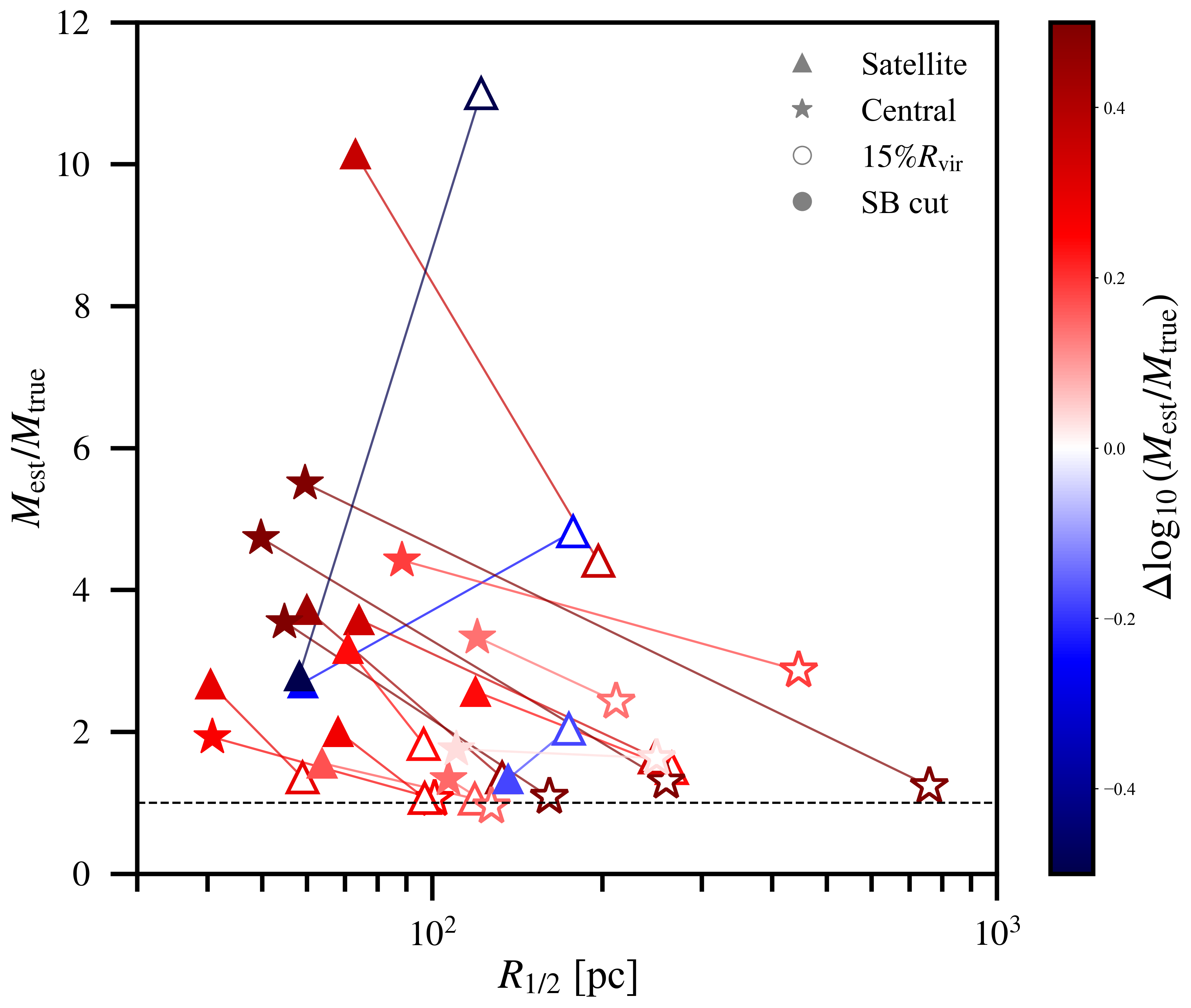} 
		\caption{ Ratio of estimated total mass enclosed within $\rhalf$ from \citet{Wolf2010}, $\mhalf^{\rm est}$ to the true total mass enclosed within the same radius, $\mhalf^{\rm true}$, for the $R_{15\%}$ (open points) and surface brightness limit ($R_{\rm SB}$; closed points). Marker shapes indicate centrals (stars) and satellites (triangles). The dashed line shows a ratio of 1, where the estimate most accurately represents the true enclosed mass. With just a few notable exceptions, removing the stellar halo from these galaxies ($R_{\rm SB}$ cut) moves the ratio to larger values and farther from 1 (i.e. the mass estimate gets worse). This is a result of the decreased true mass enclosed within a small radius rather than an increase in the \citet{Wolf2010} estimate.
		} 
        \label{fig:mratio}
	\end{figure}
        %%%%%%%%%%%%%%%%%%%%%%%%%%%%%%%%%%%%%%%%%%%%%%%%%%%%%%%%%%%%%% 
        	
% %%%%%%%%%%%%%%%%%%%%%%%%%%%%%%%%%%%%%%%%%%%%%%%%%%%%%%%%%%%%%%
% \begin{figure*}
%     \centering
%     \includegraphics[width=0.48\textwidth]{Rhalf_v_Mhalf_True_Only_10parts_Clean.png}
%     \includegraphics[width=0.48\textwidth]{Rhalf_v_Mhalf_Wolf_10parts_Clean_lines.png}
%     \caption{
%     \textbf{Left:} 2D stellar half-mass radius ($\rhalf$) versus true dynamical mass enclosed within each 3D de-projected half-light radius ($M_{1/2}^{\rm true}$) for simulated FIRE-2 dwarfs using the $R_{\rm 15\%}$ (open symbols) and $R_{\rm SB}$ (solid symbols) cuts. Marker shapes indicate centrals (stars) and satellites (triangles). Color indicates displacement between the two estimates in log-log space.
%     \textbf{Right:} Dynamical mass estimated from the Wolf et al. (2010) formula versus projected half-light radius, using the same two aperture definitions. Observed values for Milky Way and Andromeda dwarfs are shown in black with error bars \citep{Collins2013, Kirby2014, McConnachie2012}. Simulated dwarfs that meet surface brightness detection limits match observed galaxies well in both mass and size.
%     }
%     \label{fig:rhalf_versus_mhalf_both}
% \end{figure*}
% %%%%%%%%%%%%%%%%%%%%%%%%%%%%%%%%%%%%%%%%%%%%%%%%%%%%%%%%%%%%%%

%%%%%%%%%%%%%%%%%%%%%%%%%%%%%%%%%%%%%%%%%%%%%%%%%%%%%%%%%%%%%%
\begin{figure*}
    \centering
    \includegraphics[width=0.48\textwidth]{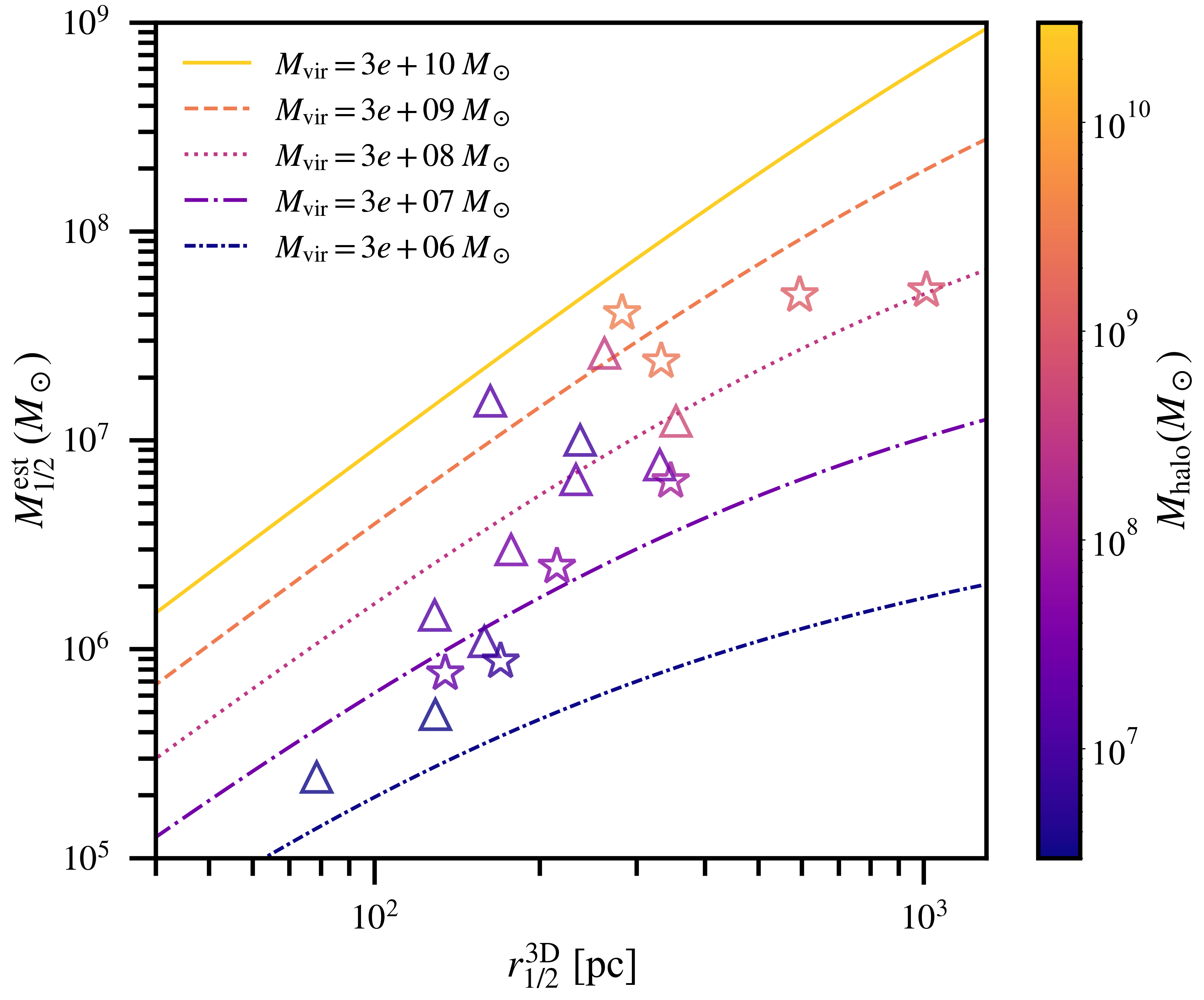}
    \includegraphics[width=0.48\textwidth]{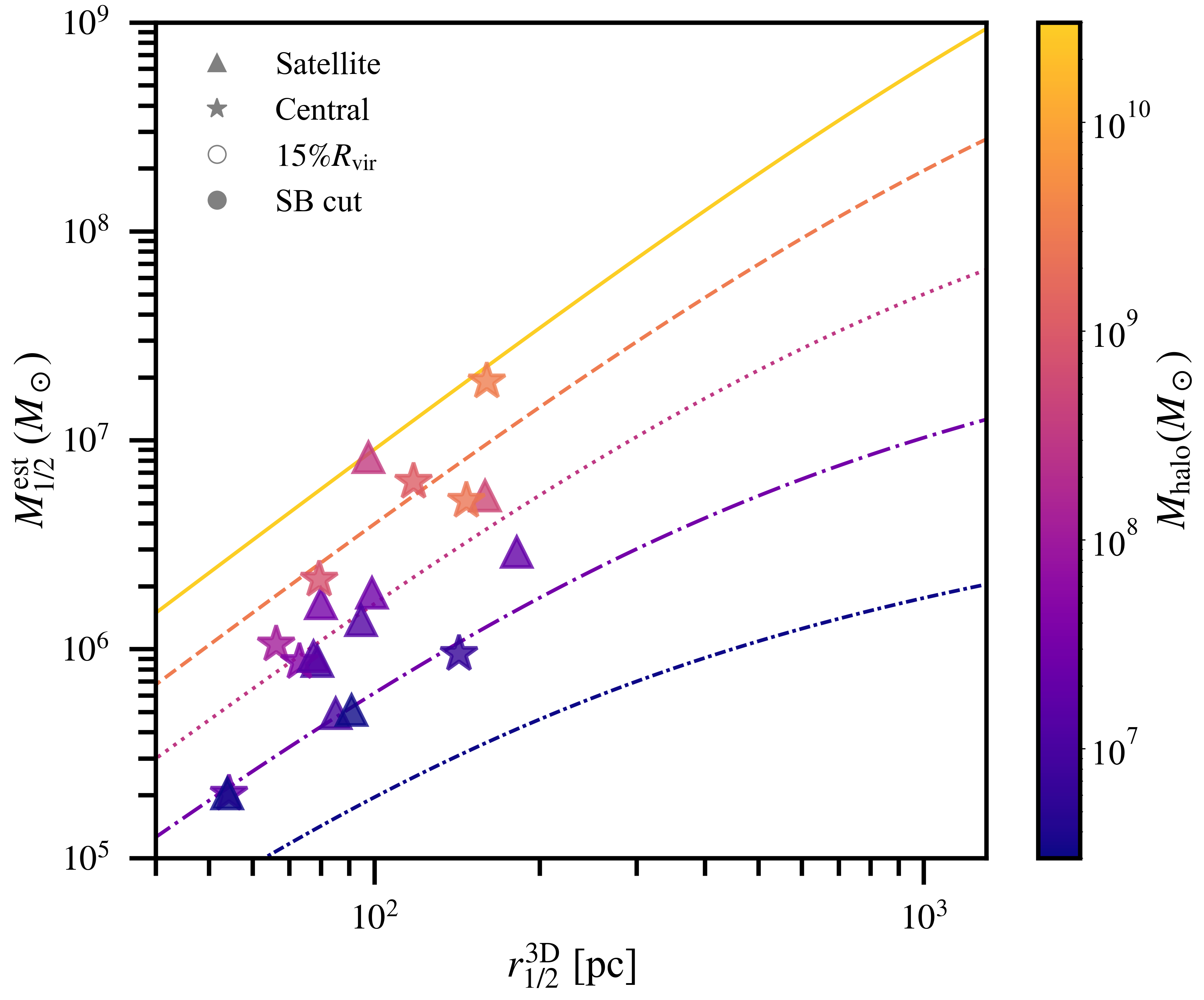}
    \caption{
    \textbf{Left:} Estimated mass within $\rhalf$ from the \citet{Wolf2010} formula, $\mhalf^{\rm est}$, versus 3D de-projected half-light radius for for simulated FIRE-2 UFDs using the $R_{\rm 15\%}$ cut. Marker shapes indicate centrals (stars) and satellites (triangles). Point color indicates the true dark matter mass, $\mhalo$, as measured by the halo finder. The lines represent the enclosed mass for NFW halo mass profiles of varying masses, color-coded the same as the points.
    \textbf{Right:} Same as in the left panel but for the $R_{\rm SB}$ cut. We assume a halo concentration of $c=24$, as predicted by extrapolating the \citet{Neto2007} halo-concentration relation down to UFD masses.
    }
    \label{fig:m_enclosed}
\end{figure*}
%%%%%%%%%%%%%%%%%%%%%%%%%%%%%%%%%%%%%%%%%%%%%%%%%%%%%%%%%%%%%%

 The \citet{Wolf2010} mass estimator has been shown to predict the masses of dynamically relaxed FIRE-2 galaxies with $\mstar\sim10^6~\msun$ to within $\sim 18\%$ \citep{Gonzalez-Samaniego2017}. 
Its accuracy relies on the assumption that the de-projected half-light radius is close to the radius $r_{-3}$ where the log-slope of the 3D tracer population is $-3$ \citep[see][]{LB2020}. This tends to be true for massive dSphs that are well-fit by Plummer or Sersic profiles, but it is not always true for more power-law like stellar mass profiles of the kind we see for our lower mass galaxies  in Fig. 1.  The problem with associating $r_{-3}$ with the half light radius is exacerbated when the outer regions are missed due to surface brightness limitations.
%
%%{\bf JSB: note that the Wolf mass is true for {\it any} velocity dispersion anisotropy -- so that's not why it doesn't work.}
%
%Spherical symmetry may be a bad assumption for UFDs \citep{Orkney2023}, and 
Indeed we see larger inaccuracies  than \citet{Gonzalez-Samaniego2017} find in higher-mass galaxies. Nevertheless, the Wolf estimator provides a handy scaling variable ($M(r) \propto \sigma^2/r$) that is often used and that we can test with and without the low-surface brightness stellar halos included.

Fig.
\ref{fig:mratio} shows $M^{\rm est}_{\rm half}/M^{\rm true}_{\rm half}$ for galaxies that include extended stellar halos, $R_{\rm 15\%}$ (open symbols), and those with galaxy edges defined by a surface brightness cut, $R_{\rm SB}$ (closed symbols). The points are color-coded by the change in the log of the mass ratio between the $R_{15\%}$ and $R_{\rm SB}$ cuts, and a line runs between methods for each galaxy. Red points represent galaxies for which $M^{\rm est}_{\rm half}/M^{\rm true}_{\rm half}$ is larger for $R_{\rm SB}$, and blue points are those where $R_{15\%}$ yields a higher ratio of the estimated to true mass. With just three exceptions, using the surface brightness cut leads to estimated mass-to-true mass ratios farther from unity than the estimate with the traditional cut often employed in simulations. Using the $R_{15\%}$ cut, the median ratio of Wolf-estimated to true enclosed mass is 1.48, whereas the ratio rises to 2.79 when using $R_{\rm SB}$. The means (2.35 for  $R_{15\%}$ and 3.31 for $R_{\rm SB}$) indicate that the surface brightness cut leads to a broader distribution of ratios, including more outliers with significantly inflated mass estimates. 

It is clear from Fig. \ref{fig:mratio} that failing to observe the stars in the stellar halo leads to a significant \textit{overestimate} of the mass enclosed within the half-mass radius, $\mhalf^{\rm est}$, compared to the true mass, $\mhalf^{\rm true}$, enclosed within the method-dependent $\rhalf$. Although  $M^{\rm est}_{\rm half}/M^{\rm true}_{\rm half}$ is higher using $R_{\rm SB}$ in all but three cases, this does not necessarily mean that the estimated mass is higher when excluding the stellar halo. When imposing the SB cut, not only does the half-mass radius decrease, the total mass within that radius, $\mhalf^{\rm true}$, \textit{decreases}, simply due to the smaller radius. This means that, if the observed ultra-faints all have extended stellar halos, and if we are underestimating their half-mass radii, we would also \textit{expect} to be measuring lower masses for those galaxies within their half-mass radii. $\mhalf^{\rm est}$ goes down, but not enough to reflect the reality of the decreased $\mhalf^{\rm true}$.

We investigate the effect of incorrectly determining $\mhalf^{\rm est}$ on the predicted halo masses in Fig. \ref{fig:m_enclosed}, where we have plotted analytic dark matter halo mass profiles with halo masses ranging from $3\times 10^{6-10}~\msun$, using the functional form from \citet[NFW; ][]{Navarro1997} and a halo concentration of $c=24$, following the extrapolated halo-concentration relation of \citet{Neto2007} to the ultra-faint mass regime studied here. Applying the $R_{\rm SB}$ cut moves most galaxies, particularly the higher mass centrals, farther to the left on this Figure than down, which leads to mass estimates that can be an order of magnitude too high. The four highest-mass centrals have halo masses $\approx 1\times 10^8~\msun-3\times10^9~\msun$. On the left, they appear to live in halos of about that mass. However, using the $R_{\rm SB}$ cut on the right places them in halos of mass $\approx 1\times10^9~\msun - 3\times 10^{10}~\msun$! Overestimating the dark matter halo mass of observed UFDs could have wide-ranging implications, including potentially indicating problems in \lcdm~ where there are none \citep{Ferrero2012}. Predicting artificially high mass-to-light ratios could warp our understanding of the low-mass end of the dark matter halo mass function and skew abundance matching results, and generally make comparison with simulations less reliable (see \citealt{Bullock2017} for a review). Furthermore, incorrect dark matter mass estimates may lead to unreliable constraints on warm, self-interacting, or fuzzy dark matter models \citep{Lovell2014,Zeng2024,Benito2025}, or on models used to test core formation methods in \lcdm ~ \citep{DiCintio2014a,Mostow2024,Orkney2021}.

	\section{Discussion and Conclusion}
	\label{sec:conclusion}

\subsection{Comparison to other work}

\citet{Klein2024} use FIRE-2 simulations of galaxies more massive ($5\times10^5\lesssim \mstar/\msun \lesssim 7\times 10^8$) than those we study here ($\mstar <10^5\msun$) to compare half-light radii in simulations using a particle-based approach and an integrated light-based approach. First limiting the galaxy edge to $10\% \rvir$, they simply sum the mass within $R_{\rm proj}$ until they reach $50\%$. They compare this $\rhalf$ to one derived from mock images using \textsc{FIRE studio}, a radiative transfer and image-generation tool developed for the \textsc{FIRE} simulations that produces mock stellar light and dust emission maps across multiple wavelengths. They find that their mock image $\rhalf$ and $\mstar$ values are each about $30\%$ lower than those derived from particle-based methods, which are more comparable to our $R_{\rm 15\%}$ model. They do employ a surface-brightness cut in their isophote at $30~\rm mag ~arcsec^{-2}$, but this method is not comparable to our $R_{\rm SB}$ cut, as we still employ a particle-based determination of the \textit{effective} surface brightness. They compare their results to the ELVES volume-limited survey of the low mass satellites of Milky Way (MW)-like hosts in the Local Volume \citep{Carlsten2022}. Using the mock image-based approach, they recover the smaller sizes and higher scatter in the mass-size plane. The primary reason for the smaller sizes using mock isophotes is that the stellar mass-to-light ratio, $\mstar/L$, is not constant with projected radius, increasing at the galaxy edges, and therefore leading to less light at larger radius. 

Fitting surface-brightness profiles to both particle distributions and to mock images is worth pursuing, and is the subject of future work (Thong et al.\ \textit{in prep}). However, the galaxy masses we study here (UFDs; $\mstar < 10^5~\msun$) are different both in that, observationally, they are detected with individual star counts, rather than with integrated light. Furthermore, they have very shallow to nonexistent age gradients, being uniformly old, and so would likely not be as impacted by gradients in the $\mstar/L$ ratio. We therefore speculate that the comparisons performed here provide a reasonably good approximation of reality, where distant stars at large radii and low effective SB are missed more easily. The recent detection of previously undetected stars at large radii lends credence to this argument.

The Engineering Dwarfs at Galaxy Formation's Edge (EDGE) simulation suite \citep{Agertz2019} contains a sample of 10 objects, including versions of 5 unique halos in the sample that have been modified to form stars earlier or later, at slightly higher mass ($5\times10^4 \lesssim \mstar/\msun \lesssim 2\times 10^ 6$) than we consider here \citep{Rey2019,Goater2024}. Investigating the formation of extended stellar structures, \citet{Rey2019} find that ``genetically modifying" their lowest-mass UFD to form later increases the fraction of stars in the galaxy that form \textit{ex-situ} and become a part of the final galaxy through late-time dry mergers. This in turn causes the galaxy to both have a larger $\rhalf$ and a lower $\mstar$ than their earlier-forming scenario, in which most of the stellar population forms from \textit{in situ} star formation. \citet{Goater2024} find that these late-time accretion events further serve to increase the ellipticity of galaxies, particularly when the lower SB features are included in the calculation. They show that this increased ellipticity is seen in their tidally isolated sample, suggesting that the extended stellar structures recently found around observed MW satellites, and their low surface brightnesses, may not necessarily be caused by tidal interactions as previously thought \citep{Li2018,Munoz2010}. Given that our simulations also do not have a Milky Way-mass host galaxy, our results suggest that large sizes do not require tides \citep{Mercado2025}. All of our extremely low-mass UFDs form stars within the first $500 \rm ~Myr$ of the simulation, and so their puffy nature is more likely related to bursty star formation and late-time accretion events.

\citet{Ko2024} conduct high-resolution ($\mbar =60~\msun$) simulations of six UFD analogs, focusing on the effect of their galaxies assembling from multiple progenitors and how stellar mass, metals and galaxy sizes build up over time. The halo that forms from a single progenitor achieves $\rm [Fe/H]= -2.19$, This supports the results of \citet{Goater2024} in which 5 galaxies, with stellar masses comparable to the most massive twon objects we study here ($\mtenvB$ and $\mnine$), are an entire dex closer to observations, even when including the lowest metallicity ($\feh <-4$) stars. While they credit their discrete sampling of the initial mass function (IMF) for the higher metallicity, the larger differences between this work and theirs lies in our lack of any model for pop III star formation (other than the metallicity floor), and their sole inclusion of thermal rather than kinetic core-collapse SN feedback (we include both). Interestingly, \citet{Ko2024} find that their simulations have many stars out to 10 half-light radii, and are better fit with a two-component density profile than a single elliptical exponential. When they plot the effective radii for the two-component fit versus the total luminosity, they achieve good agreement with observations. Using the single-component profile, one of their galaxies does reach $\rhalf > 1\rm ~kpc$, even larger than our results. This suggests that two-component profiles may be a better way to determine half-light radii in UFDs. We explore this idea in future work (Murphy et al. \textit{in prep}).

\subsection{Summary of Results}

We investigate the effect of redefining a simulated galaxy ``edge" via a surface brightness cut -- effectively removing their extended stellar halos -- on the properties of 19 ultra-faint galaxies in a suite of $\mbar = 30\,M_{\odot}$ cosmological zoom-in FIRE-2 simulations first described in \citet{Wheeler2019}. We apply both a cut at $15\%$ of the virial radius of each galaxy's dark matter halo, $R_{\rm 15\%}$, and a mock observational surface brightness threshold at $\mu_V \approx 32.5$ mag arcsec$^{-2}$, $R_{\rm SB}$. With these two definitions for defining the extent of a galaxy, we compare predictions with each galaxy edge scheme to each other and to a compilation of observed UFDs ($\mstar < 10^5~\msun$) from \citet{Pace2024}. We summarize our main conclusions below: 
\begin{itemize}

\item We implement a galaxy edge cut at an effective surface brightness of $32.5 \rm ~mag~arcsec^{-2}$ and compare predictions for UFD properties to those made using a fixed fraction of the virial radius, as is common in simulations, specifically $15\%~\rvir$. Using $R_{\rm SB}$ instead of $R_{\rm 15\%}$  decreases the stellar mass and reduces the inferred size of ultra-faint galaxies ($\rhalf$). This moves predictions for simulated FIRE-2 galaxies \textit{closer to the observed UFD population on the mass-size plane,} and closer to effective surface brightness limits of current surveys such as SDSS.

\item The same cut, $R_{\rm SB}$, causes galaxies to exhibit slightly lower velocity dispersions ($\sigma_{\rm los}$) which, when combined with the reduced $\mstar$ values, shifts predictions towards the extreme low-mass end of observed values in the $\sigma_{\rm los}-\mstar$ plane but ultimately is \textit{more consistent with observations} than the $R_{15\%}$ cut. Using the surface brightness restriction pushes most velocity dispersion values below $5~\kms$ without the need for tidal stripping from a massive galaxy. There is a trend between $\sigma_{\rm los}$ and $\mhalo$ for $R_{\rm SB}$, and a much weaker one for the $R_{15\%}$ due to several high velocity dispersion outliers. Central galaxies with $\sigma_{\rm los}<7~\kms$ exhibit a positive correlation between halo mass and surface brightness (lower $\mhalo$ means higher $\mu$ value and lower surface brightness) when using the $R_{\rm SB}$ cut, as predicted by \citet{Bullock2010}, but exhibit the opposite trend using the $R_{\rm 15\%}$ cut.

\item  We find that the SB cut mildly alleviates the well-known discrepancy between simulated and observed UFDs on the MZR for the two most massive UFDs ($\mnine$ and $\mtenvB$), suggesting that \textit{observed [Fe/H] values may be  biased slightly high due to missing low metallicity stars at large radii.}  The same cut has minimal impact on lower-mass UFDs, likely due to the absence of metallicity gradients in the latter. We note that removing the large unenriched tail of the two most massive UFDs, as may be done by assuming the stars are pop III stars, would also alleviate the problem, moving the mean [Fe/H] to $-3$ for both models.

\item Comparing the \citet{Wolf2010} mass estimator to the circular velocity curves calculated from the true total enclosed mass in the simulations indicates that \textit{surface brightness cuts bring predictions closer to observations,} but can cause overestimation of galaxy masses, or at least an overestimate of the halo density. This is consistent with calculations of the ratio of $\mhalf^{\rm est}$ to $\mhalf^{\rm true}$, which we show to move further from unity after imposing the $R_{\rm SB}$ cut. 
%However, this is the result of \textit{decreased} true mass enclosed within the new $\rhalf$ values rather than an actual increase in the \citet{Wolf2010} mass estimator.
Both true mass and estimated mass decrease with the $R_{\rm SB}$ cut, but true mass more so.

\item We emphasize our primary result, \textit{excluding the stellar halo, i.e. by using the $R_{\rm SB}$ cut, moves predictions for galaxy size and $\mhalf^{\rm est}$ closer to observations, while simultaneously moving $\mhalf^{est}$ farther from the true enclosed mass, $\mhalf^{\rm true}$.} Removing the stars at the outskirts of ultra-faint galaxies can incorrectly increase their dark matter halo mass estimates by an order of magnitude. This suggests that, not only is it possible that we are missing extended stellar halos around many UFDs, but that current mass estimates could be significantly over-predicting the density of the host DM halos at the lowest masses.

	\end{itemize}

	\section*{Acknowledgments} 
   CW acknowledges the CPP Provost's Teacher-Scholar Program. MKRW acknowledges support from NSF MPS Ascending Faculty Catalyst Award AST-2444751. FJM is funded by the National Science Foundation (NSF) Award AST-2316748. JSB is supported by NSF grant AST-2408247 and NASA grant 80NSSC22K0827. MBK acknowledges support from NSF grants AST-1910346, AST-2108962, and AST-2408247; NASA grant 80NSSC22K0827; HST-GO-16686, HST-AR-17028, HST-AR-17043, JWST-GO-03788, and JWST-AR-06278 from the Space Telescope Science Institute, which is operated by AURA, Inc., under NASA contract NAS5-26555; and from the Samuel T. and Fern Yanagisawa Regents Professorship in Astronomy at UT Austin. SRL acknowledges support from NSF grant AST-2109234 and HST grant AR-16624 from STScI. All authors would like to thank all of the amazing the people who make our work possible: the cleaning and the clerical staff, the food service workers, the technical support personnel, and so many more.
   
\bibliographystyle{aasjournal}
\bibliography{dws.bib}

\end{document}